\documentclass[fleqn,usenatbib]{mnras}

\usepackage{mathptmx}
\usepackage{txfonts}

\usepackage[T1]{fontenc}

\DeclareRobustCommand{\VAN}[3]{#2}
\let\VANthebibliography\thebibliography
\def\thebibliography{\DeclareRobustCommand{\VAN}[3]{##3}\VANthebibliography}

\usepackage[switch]{lineno}
\usepackage{txfonts}
\usepackage{natbib}
\usepackage{graphicx, xspace}
\usepackage{multirow}
\usepackage{amssymb}
\usepackage{subfigure}
\usepackage{epstopdf}
\usepackage{lineno}
\usepackage{cancel}

\title[Earth's habitability over the Sun's main-sequence]{On Earth's habitability over the Sun's main-sequence history: joint influence of space weather and Earth´s magnetic field evolution}

\author[J. Varela et al.]{
J. Varela,$^{1}$\thanks{E-mail: jvrodrig@fis.uc3m.es}
A. S. Brun,$^{2}$
A. Strugarek,$^{2}$
V. R\'eville,$^{3}$
P. Zarka$^{4}$
and F. Pantellini$^{5}$
\\
$^{1}$Universidad Carlos III de Madrid, Leganes, 28911\\
$^{2}$Laboratoire AIM, CEA/DRF – CNRS – Univ. Paris Diderot – IRFU/DAp, Paris-Saclay, Gif-sur-Yvette Cedex, France, 91191\\
$^{3}$IRAP, Universit\'e Toulouse III—Paul Sabatier, CNRS, CNES, Toulouse, France\\
$^{4}$LESIA \& USN, Observatoire de Paris, CNRS, PSL/SU/UPMC/UPD/UO, Place J. Janssen, Meudon, France, 92195\\
$^{5}$LESIA, Observatoire de Paris, Universit\'e PSL, CNRS, Sorbonne Universit\'e, Universit\'e de Paris, 5 place Jules Janssen, Meudon, France, 92195 
}

\date{Accepted XXX. Received YYY; in original form ZZZ}

\pubyear{2023}

\begin{document}
\label{firstpage}
\pagerange{\pageref{firstpage}--\pageref{lastpage}}
\maketitle

\begin{abstract}
The aim of this study is to analyze the Earth habitability with respect to the direct exposition of the Earth atmosphere to the solar wind along the Sun’s evolution on the main sequence including the realistic evolution of the space weather conditions and the Earth magnetic ﬁeld. The MHD code PLUTO in spherical coordinates is applied to perform parametric studies with respect to the solar wind dynamic pressure and the interplanetary magnetic ﬁeld intensity for diﬀerent Earth magnetic ﬁeld conﬁgurations. Quiet space weather conditions may not impact the Earth habitability. On the other hand, the impact of interplanetary coronal mass ejections (ICME) could lead to the erosion of the primary Earth atmosphere during the Hadean eon. A dipolar ﬁeld of $30$ $\mu$T is strong enough to shield the Earth from the Eo-Archean age as well as $15$ and $5$ $\mu$T dipolar ﬁelds from the Meso-Archean and Meso-Proterozoic, respectively. Multipolar weak ﬁeld period during the Meso-Proterozoic age may not be a threat for ICME-like space weather conditions if the ﬁeld intensity is at least $15$ $\mu$T and the ratio between the quadrupolar ($Q$) and dipolar ($D$) coeﬃcients is $\frac{Q}{D} \leq 0.5$. By contrast, the Earth habitability in the Phanerozoic eon (including the present time) can be hampered during multipolar low ﬁeld periods with a strength of $5$ $\mu$T and $\frac{Q}{D} \geq 0.5$ associated to geomagnetic reversals. Consequently, the eﬀect of the solar wind should be considered as a possible driver of Earth´s habitability.
\end{abstract}




\section{Introduction}

The analysis of the space weather in the last decades has demonstrated the essential role of the solar wind (SW) and interplanetary magnetic field (IMF) on the Earth magnetosphere, ionosphere, thermosphere and exosphere state \citep{Poppe,Gonzalez}. That means, space weather conditions can introduce constraints on the habitability of the Earth and exoplanets with respect to the shielding provided by planetary magnetospheres, avoiding the sterilizing effect of the stellar wind on the surface \citep{Gallet,Linsky,Airapetian,Strugarek,Garraffo,Varela7}. Recent studies also indicate the solar wind may cause the depletion of the Earth atmosphere, particularly volatile molecules such as water by thermal and non-thermal escape \citep{Lundin,Moore,Jakosky} as well as lead to damages on biological structures \citep{Sihver,Zenchenko}.

The largest perturbations of the Earth magnetosphere is observed during periods of extreme space weather conditions associated to coronal mass ejections (CME) \citep{Cane,Richardson,Wang4,Lugaz,Wu}. The CMEs are solar eruptions caused by magnetic reconnections in the star corona \citep{Low,Howard}, leading to the generation of streams of charged particles and a magnetic cloud that propagates in the interplanetary medium (ICME) \citep{Sheeley,Neugebauer,Cane2,Gosling}. During the impact of an ICME with the Earth, the SW dynamic pressure and the intensity of the IMF can increase by two orders of magnitude \citep{Gosling2,Huttunen,Manchester,Schwenn,Riley,Howard2,Mays,Kay,Savani,Salman,Kilpua,Hapgood,Regnault,Regnault2}. Damages in spacecrafts \citep{Choi,Nwankwo,Molera} and critical infrastructures such as electric power grids \citep{Cannon} are observed during ICME. In addition, ICMEs may hamper the habitability of exoplanets hosted by stars showing a large magnetic activity (larger ICME recurrence and intensity than the Sun) \citep{Khodachenko,Lammer,Balona,Schaefer}.

Early stages of the Sun evolution showed a fast rotation rate and high magnetic activity \citep{Pizzolato,Emeriau} that decreased along the Sun's evolution on the main sequence \citep{Folsom,Fabbian}, a mechanism known as gyro-chronology \citep{Skumanich,Barnes}. Accordingly, the forcing of the SW and IMF on the Earth magnetosphere, that is to say the space weather conditions, also changed \citep{Reville,Carolan,Ahuir,Varela7}. The SW dynamic pressure and IMF intensity were much higher at early stages of the Sun main sequence compared to the present days \citep{Airapetian4,Lugaz3}, thus the perturbations induced by the young Sun on the Earth magnetosphere were stronger \citep{Sterenborg,Carolan,Varela7}, and the Earth likely exposed to more powerful and recurrent ICMEs \citep{Sterenborg,Airapetian2,Airapetian3,Shibayama} leading to an increase of the ion transport in the inner magnetosphere \citep{Zhang4}.

Geological surveys \citep{Schwarz,McElhinny,Tarduo,Tarduo2,Tarduo3,Tarduo4} and geodynamo numerical models \citep{Aubert,Aubert2,Driscoll,Davies,Aubert3} indicate a large variability of the Earth magnetic field intensity and topology along the Sundas evolution on the main sequence. There are evidences of weak field periods during the Paleo-Archean and Meso-Archean eras as well as the Proterozoic eon showing dipolar field intensities below $20$ $\mu$T \citep{Morimoto,Elming,Chiara}. In addition, weak field periods are also identified along the Paleozoic (Cambrian, Denovian and Carboniferous periods) \citep{Hawkins,Hawkins2,Thallner} and Mesozoic eras (Triassic, Jurassic and Paleogene periods) \citep{Prevot,Kosterov,Juarez,Heunemann,Rathert}. On the other hand, strong dipolar field periods could exist during the Hadean eon and Eo-Archean era showing intensities above $40$ $\mu$T \citep{Hale,Tarduo4}. Nevertheless, periods of dipolar field with intensities comparable to the present time were also held in the Meso-Proterozoic and Neo-Archean eras as well as in the Cretaceous and Neogene periods \citep{Yoshihara,Chiara2,Kaya}. In addition, there are evidences of multipolar low field configurations during the Proterozoic age, the Devonian and Cretaceous periods linked to geomagnetic reversals \citep{McElhinny,Bogue,Gubbins}, characterized by large quadrupolar components of the magnetic field similar to the present Hermean magnetic field \citep{Lhuillier,Shcherbakova,Guyodo}.

The aim of the present study is to analyze the Earth habitability along the Sun's evolution on the main sequence from the point of view of the space weather conditions and Earth magnetic field configuration, that is to say, identifying if the Earth magnetosphere was able to protect the planet from the direct impact of the SW in the last $4600$ million years. In addition, the study explores the minimum requirements of the magnetic field in the next $3000$ million years to protect the Earth (Posterum eons). The study methodology is based on the extensive use of parametric analysis using a global MHD model, calculating the magnetosphere deformation for different space weather conditions and configurations of the Earth magnetic field. The study includes space weather conditions from quiet to super ICME events as well as dipolar and multipolar fields with intensities consistent with the paleomagnetism data. 

The analysis of the Earth habitability conditions is performed with respect to the magnetopause standoff distance, assuming that the Earth habitability could be hampered if the magnetopause collapses onto the Earth surface, that is to say, the SW precipitates directly towards the Earth. Consequently, the magnetosphere fails to protect both the Earth surface and atmosphere. It should be noted that there are other important factors that constrain the Earth habitability, for example the EUV, X ray and cosmic rays \citep{Buccino,Forcada,Airapetian5,Johnstone2,Rodgers}. In particular, large EUV and X-ray fluxes during the Hadean and Archean eons may have caused an expansion of the ionosphere affecting the standoff distance of the bow shock and magnetopause \citep{Ribas,Tu,Lammer3}. Nevertheless, the SW precipitation should be understood as one of the main concerns for the Earth habitability along the Sun's evolution on the main sequence \citep{Zendejas,See,Blackman,Mozos,Chebly}.

MHD models are routinely used to analyze the interaction of the SW and IMF with planetary magnetospheres \citep{2008Icar..195....1K,2015JGRA..120.4763J,Strugarek2,Strugarek,Varela,Paul}. Previous studies indicate a stronger compression of the bow shock as the dynamic pressure of the SW increases and a modification of the planetary magnetosphere topology caused by the interaction between the IMF and the planetary magnetic field \citep{Slavin,2000Icar..143..397K,2009Sci...324..606S,Varela4}. MHD models are also applied to study the effect of the space weather on the Earth magnetosphere's large scale structures, for example the Bow Shock \citep{Samsonov,Andreeova,Nemecek,Mejnertsen}, the Magnetosheath \citep{Ogino,Wang}, the magnetopause stand off distance \citep{Cairns,Cairns2,Wang2} and the magnetotail \citep{Laitinen,Wang3}. In particular, global MHD models reproduce the strong deformations induced by ICMEs in the Earth magnetosphere \citep{Wu2,Wu3,Shen,Ngwira,Wu4,Scolini,Torok} leading to an important decrease of the magnetopause stand off distance \citep{Sibeck,Dusik,Liu,Nemecek2,Grygorov,Samsonov2}. It should be noted that MHD codes were validated comparing the simulation results with ground based magnetometers and spacecraft measurements \citep{Raeder2,Wang6,Den,Facsko,Honkonen}. Nevertheless, kinetic effects are also important but are out of the scope of the present study and may have effects on the final conclusions \citep{Xiaojun,Omelchenko,Lembege}.

The analysis is performed using the single fluid MHD code PLUTO in spherical 3D coordinates \citep{Mignone}. The numerical framework was already applied to model global structures of the Hermean magnetosphere \citep{Varela,Varela2,Varela3,Varela4}, the effect of ICME-like space weather conditions on the Earth magnetosphere \citep{Varela7}, slow modes in the Hermean magnetosphere \citep{Varela8} as well as the radio emission from the Hemean and exoplanetary magnetospheres \citep{Varela5,Varela6,Varela9}.

This paper is structured as follows. There is a basic description of the simulation model in section \ref{Model}. The space weather models used in the analysis are discussed in section \ref{weather}. The Earth habitability from the Hadean eon to the present era with respect to the intensity of a dipolar magnetic field is analyzed in section \ref{Dipole}. The consequences on the Earth habitability of periods with multipolar magnetic fields are studied in section \ref{Multipolar}. The minimum Earth habitability requirements for a decaying Earth magnetic field in future eons is analyzed in section \ref{Posterum}. The effect of an expanded ionosphere due to large EUV and X-ray fluxes on the Earth habitability is introduced in section \ref{Ionosphere}. Finally, section \ref{Conclusions} shows the summary of the study main conclusions discussed in the context of other authors results.

\section{Numerical model}
\label{Model}

The analysis is performed using the ideal MHD version of the open-source code PLUTO in spherical coordinates. The model calculates the evolution of a single-fluid ideal plasma in the nonresistive and inviscid limit \citep{Mignone}. A detailed description of the model equations, boundary conditions and upper ionosphere model can be found in \cite{Varela7}.

The simulations are performed using a grid of 128 radial points, 48 bins in the polar angle $\theta$ and 96 in the azimuthal angle $\phi$, equidistant in the radial direction. The simulation domain is confined between two concentric shells around the Earth, an outer boundary ($R_{out}$) fixed at $30$ $R_{E}$ ($R_{E}$ is the Earth radius) and an inner boundary ($R_{in}$) whose radial location depends on the intensity of the Earth magnetic field. In addition, below the inner boundary of the simulation domain, an upper ionosphere coupling region is added based on the electric field generated by the field-aligned currents providing the plasma velocity at the upper ionosphere \citep{Buchner}. The radial extension of the upper ionosphere coupling region ($\Delta R_{ion}$) is fixed to $0.2$ $R_{E}$. The effect of the ionosphere expansion caused by large EUV and X-ray fluxes in the Hadean and Archean eons \citep{Ribas,Tu,Lammer3} is introduced in the model. This first attempt to improve the ionosphere model is discussed in the Appendix section. Such extended model is not used in all the simulation set, only in the cases studied in section \ref{Ionosphere}.

A set of Earth magnetic field configurations are considered in the analysis. Low to high dipolar magnetic field periods are analyzed using dipole models with an intensity of $5$, $15$, $30$ and $45$ $\mu$T. The intensity of the  magnetic field models is defined at the surface of the Earth in the equatorial region (ranging for the case of the present Earth between $25$ and $65$ $\mu$T between the equator and the poles \citep{Finlay}). Multipolar low field periods are explored by models including dipolar and quadrupolar field components, parameterized by the ratio between the quadrupole ($Q$) and dipole ($D$) coefficients, $Q/D$, and the field intensity. Particularly, the simulations are performed for the models with $Q/D = 0.2$ and $0.5$ as well as field intensity of $5$ and $15$ $\mu$T (the combination of dipolar and quadrupolar components of the magnetic field leads to the given intensity at the Earth surface in the equatorial region). Future scenarios are simulated as a decaying dipolar field with an intensity of $0.1$ and $1$ $\mu$T (geo-dynamo extinction). Figure \ref{1} summarizes the Paleomagnetic data obtained from geological surveys and the Earth magnetic field model applied to study different geological times.

\begin{figure*}
\centering
\includegraphics[width=15cm]{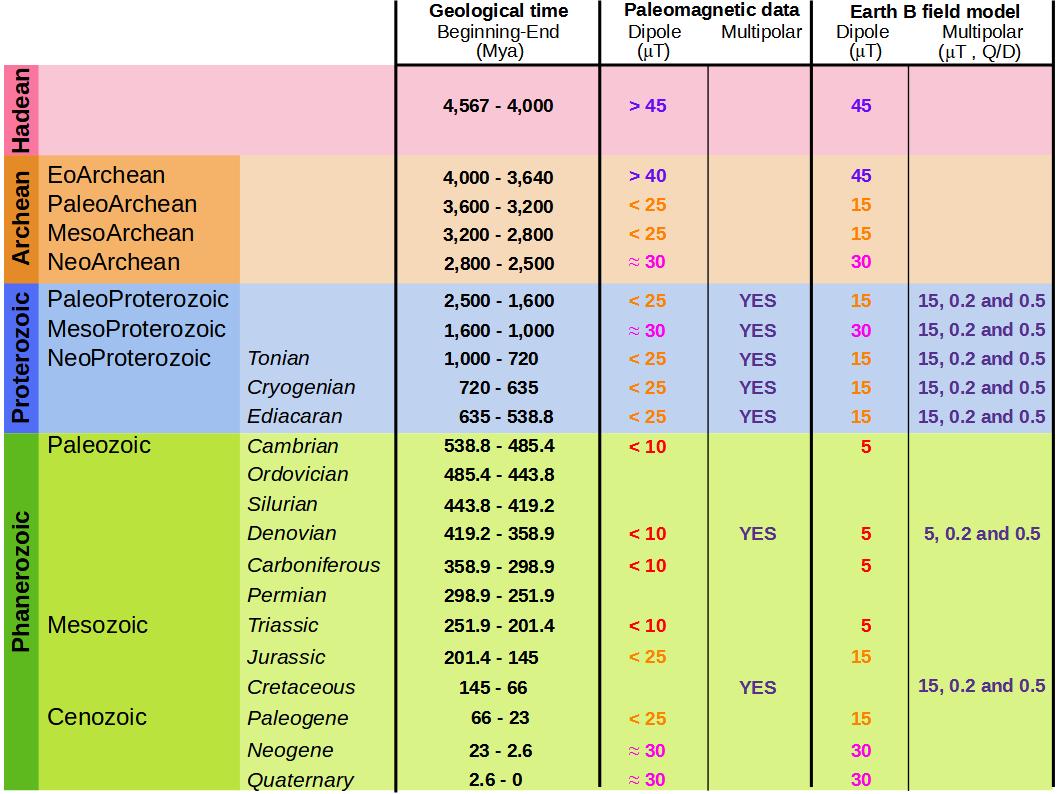}
\caption{Earth magnetic field intensity (first column) and evidence of large multipolar components (second column) obtained from Paleomagnetic data. Earth magnetic field model applied to study geological times with a dominant dipolar component (third column) or multipolar configurations (fourth column).}
\label{1}
\end{figure*}

The Earth magnetic field is rotated $90^{o}$ in the YZ plane with respect to the grid poles, a transformation required to avoid spurious plasma outflows in the cusp regions linked to an issue with the boundary condition where geometrical and magnetical poles overlap (no special treatment was included for the singularity at the magnetic poles). Consequently, the IMF orientation is adapted accordingly. The tilt of the Earth rotation axis is included (fixed at $23^{o}$ with respect to the ecliptic plane). The effect of the Earth rotation is not included in the analysis for simplicity.

The simulation frame assumed is: z-axis is provided by the planetary magnetic axis pointing to the magnetic north pole, star-planet line is located in the XZ plane with $x_{star} > 0$ (solar magnetic coordinates) and the y-axis completes the right-handed system.

The set of simulations performed for different space weather conditions are parameterized with respect to the SW dynamic pressure and IMF intensity. A Southward IMF is assumed because it is the orientation leading to the strongest reconnection between the IMF and Earth magnetic field and the lowest magnetopause standoff distance. Consequently, the analysis provides the upper limit of the SW restrictions on the Earth habitability for a given space weather configuration. The decayed dipole simulations also include in the analysis the Northward, radial (Earth-Sun) and ecliptic (counter-clock wise) IMF orientations. It should be noted that the space weather conditions leading to a magnetopause standoff distance below the inner boundary of the simulation domain (the top of the upper ionosphere model, not the planet surface) or numerical instabilities causing the termination of the simulation are removed from the analysis.

Table \ref{1} shows the main parameters for each Earth magnetic field model as well as the range of SW dynamic pressure and IMF intensity values explored. SW temperature is $T = 2 \cdot 10^{5}$ K in all the configurations analyzed.

\begin{table*}
\centering
\begin{tabular}{c c | c c c c}
Topology & $B_{E}$ & $R_{in}$ & $P_{d}$ & $B_{IMF}$ & Configuration \\
($Q/D$) & ($\mu$T) & ($R_{E}$) & (nPA) & (nT) & \\ \hline
$0$ & $0.1$ & $1.0$ & $1$-$5$ & $1$-$50$ & decayed dipole \\
$0$ & $1$ & $1.0$ & $1$-$100$ & $1$-$150$ & decayed dipole \\
$0$ & $5$ & $1.5$ & $1$-$200$ & $1$-$200$ & Low dipole field \\
$0$ & $15$ & $2.25$ & $1$-$200$ & $1$-$200$ & Low dipole field \\
$0$ & $30$ & $2.5$ & $1$-$200$ & $1$-$200$ & Standard dipole field \\
$0$ & $45$ & $2.5$ & $1$-$250$ & $1$-$250$ & High dipole field \\
$0.2$ & $5$ & $1.2$ & $1$-$150$ & $1$-$200$ & Multipolar \\
$0.5$ & $5$ & $1.2$ & $1$-$50$ & $1$-$150$ & Multipolar \\
$0.2$ & $15$ & $1.2$ & $1$-$200$ & $1$-$200$ & Multipolar \\
$0.5$ & $15$ & $1.2$ & $1$-$200$ & $1$-$200$ & Multipolar \\
\end{tabular}
\caption{Main simulation parameters for each Earth magnetic field model: quadrupole - dipole coefficients ratio, field intensity (at the equatorial region of the Earth surface), inner boundary location, SW dynamic pressure range, IMF intensity range, and Earth magnetic field configuration.}
\label{1}
\end{table*}

Figure \ref{2} shows a 3D view of a simulation domain for typical space weather conditions during the Cambrian period (SW dynamic pressure of $1$ nPa and a Southward IMF intensity of $10$ nT) for a dipolar weak field period of $5$ $\mu$T. The equatorial and polar cuts indicate the SW density distribution showing the formation of the bow shock in the Earth day side due to the accumulation of plasma coming from the slowed down and diverted SW by the Earth magnetosphere. The SW flow causes the bending of the Earth magnetic field lines and the compression of the magnetosphere. The Southward IMF reconnects with the Earth magnetic field and causes the erosion of the magnetosphere in the equatorial region, reducing the efficiency of the magnetic shield to protect the Earth surface and atmosphere. The magnetopause is located several Earth radii above the surface, thus the Earth magnetic field is strong enough to avoid the direct precipitation of the SW. It should be noted that the star is not included inside the computational domain and the space weather conditions are introduced as the outer boundary conditions of the model in the day side.

\begin{figure*}
\centering
\includegraphics[width=15cm]{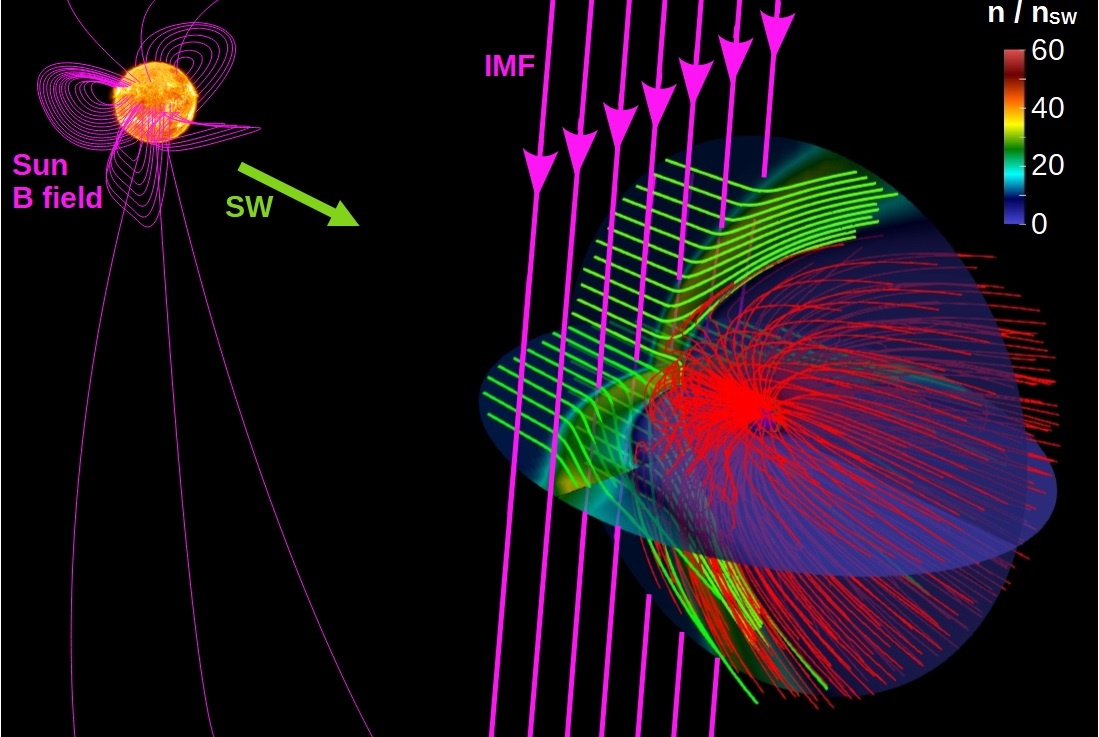}
\caption{3D view of a typical simulation setup. SW density distribution (color scale), Earth magnetic field lines (red lines), and Sun magnetic field (pink lines). The bold pink arrows indicate the orientation of the IMF (Southward orientation). The green lines indicate the SW velocity stream lines.}
\label{2}
\end{figure*}

The theoretical approximation of the magnetopause standoff distance is calculated as the balance between the dynamic pressure of the SW ($P_{d} = m_{p} n_{sw}  v_{sw}^{2}/2$), the thermal pressure of the SW ($P_{th,sw} = m_{p} n_{sw}  v_{th,sw}^{2}/2 = m_{p} n_{sw} c_{sw}^{2}/\gamma$), and the magnetic pressure of the IMF ($P_{mag,IMF} = B_{IMF}^{2}/2 \mu_{0})$ with respect to the magnetic pressure of the compressed Earth dipolar magnetic field ($P_{mag,E} = \alpha \mu_{0} M_{E}^{2} / 8 \pi^2 r^{6} $) and the thermal pressure of the magnetosphere ($P_{th,MSP} = m_{p} n_{MSP}  v_{th,MSP}^{2}/2$). Here, $m_{p}$ is the proton mass, $n_{sw}$ the SW density, $v_{sw}$ the SW velocity, $c_{sw}$ the SW sound speed, $\gamma$ the polytropic index, $B_{IMF}$ the IMF intensity, $\mu_{0}$ the vacuum magnetic permeability, $\alpha$ the dipole compression coefficient ($\alpha \approx 2$ \citep{Gombosi}), $M_{E}$ the Earth's dipole magnetic field moment, $n_{MSP}$ the magnetospheric plasma density and $v_{th,MSP}$ the thermal velocity of the magnetospheric plasma. This results in the expression:
\begin{equation}
\label{Equation B1}
P_{d} + P_{mag,sw} + P_{th,sw} = P_{mag,E} + P_{th,MSP}
\end{equation}
thus,
\begin{equation}
\label{Equation B2}
\frac{R_{mp}}{R_{E}} = \left[ \frac{\alpha \mu_{0} M_{E}^{2}}{4 \pi^2 \left( m_{p} n_{sw}  v_{sw}^{2} + \frac{B_{IMF}^{2}}{\mu_{0}} + \frac{2 m_{p} n_{sw}  c_{sw}^{2}}{\gamma} - m_{p} n_{MSP} v_{th,MSP}^{2} \right)} \right]^{(1/6)}
\end{equation}
with $r_{mp} = R_{mp} / R_{E}$ the magnetopause standoff distance. It should be noted that this expression holds for a dipolar magnetic field, not for a multipolar magnetic field. The description of a multipolar field requires introducing the effect of high order moments of the mutipolar expansion in the definition of the magnetic pressure of the Earth magnetic field. The magnetopause standoff distance in the simulations analysis is defined as the last close magnetic field line on the exoplanet dayside at $0^{o}$ longitude in the ecliptic plane. The data set of magnetopause standoff distance values calculated for different space weather conditions are fitted with respect to $P_{d}$ and $|B|_{IMF}$ using the surface function $log(r_{mp}) = log(Z) + Mlog(|B|_{IMF}) + Nlog(P_{d})$, derived from the expression $r_{mp} = Z |B|_{IMF}^{M} P_{d}^{N}$ with $Z$ a proportionality constant. The critical $|B|_{IMF}$ and $P_{d}$ values indicating the Earth habitability threshold are calculated as the isoline that verifies $r_{mp} = 1$, that is to say, direct precipitation of the SW towards the planet.

\section{Space weather modeling}
\label{weather}

The evolution of the space weather conditions generated by the Sun along the main sequence can be predicted applying different model. For example, the evolution of the SW velocity and density at the Earth orbit can be estimated using the model by \citet{Griemeier}, that combines the empirical estimation of the evolution of the stellar mass loss rate as a function of stellar age by \citet{Wood} and a model of the age-dependence of the stellar wind velocity by \citet{Newkirk}:
\begin{equation}
\label{Equation B3}
v_{sw}(t) = v_{0} \left(1+\frac{t}{\tau}\right)^{-0.43}
\end{equation}
\begin{equation}
\label{Equation B4}
n_{sw}(t) = n_{0} \left(1+\frac{t}{\tau}\right)^{-1.84 \pm 0.6}
\end{equation}
providing a first order of magnitude approximation of the $P_{d}$ values along the Sun´s evolution on the main sequence for quiet space weather conditions. ICME-like space weather conditions are defined assuming an increase of $P_{d}$ by two orders of magnitude compared to the quite conditions, consistent with the upper bound of $P_{d}$ during the observed ICMEs \citep{Gosling2,Riley,Savani,Salman,Kilpua,Hapgood}. It should be noted that super-flares may cause $P_{d}$ above the parametric range analyzed \citep{Maehara,Maehara2}, thus super-flares are out of the scope of the study. Next, the critical $B_{IMF}$ required for the direct SW deposition during quiet and ICME-like space weather conditions is calculated taking the $P_{d}$ values from \citet{Griemeier} and using the regression results imposing the condition $r_{mp}=1$.

The critical $B_{IMF}$ calculated can be compared with models that predict the IMF intensity at the Earth orbit along the Sun's evolution on the main sequence. One possibility is using the model developed by \citet{Vidotto,Carolan} that show a reasonable agreement with the SW dynamic pressure values suggested by \citet{Griemeier}. The \citet{Vidotto,Carolan} approximation is based on a $1.5D$ Weber-Davis model \citep{Weber} solved using the Versatile Advection Code (VAC) \citep{Toth,Johnstone} assuming a polytropic wind of $1.05$ leading to an almost isothermal stellar wind temperature profile. If the calculated critical $B_{IMF}$ is above the IMF intensity in \citet{Vidotto,Carolan} model for a given geological time and dipole intensity, the Earth habitability could be jeopardized by the effect of the space weather conditions. The analysis assumes the upper bound of the IMF intensity during ICMEs is $50$ times larger compared to the quiet space weather conditions provided by \citet{Vidotto,Carolan}. In the following, the the combination of \citet{Griemeier} and \citet{Vidotto,Carolan} models to calculate the space weather conditions is called G-V-C model.

In addition to the space weather predictions by the G-V-C model the analysis also includes the estimations provided by the Ahuir model \citep{Ahuir}. Applying two different models is required to validate the analysis results, that is to say, if similar consequences on the Earth habitability are obtained using both models the study conclusions may be considered robust. Ahuir model brings power-laws that describe the mass loss, magnetic field as well as the coronal density and temperature evolution for solar-like stars among the main sequence, coupled with a wind torque formulation consistent with the distribution of stellar rotation periods in open clusters \citep{Reiners,Gallet2,See3,Jardine} and the Skumanich law \citep{Skumanich,Skumanich2} (empirical relationship between the rotation rate of a star and its age that assumes the equatorial speed of a star at its equator falls as the inverse square root of the star's age). Figure \ref{3} compares the SW dynamic pressure and IMF intensity values in G-V-C and Ahuir models at different geological times. A similar prediction of the IMF intensity is observed but the Ahuir model shows a lower SW dynamic pressure before the Neo-Proterozoic and higher from the Paleozoic.

\begin{figure}
\centering
\resizebox{\hsize}{!}{\includegraphics[width=\columnwidth]{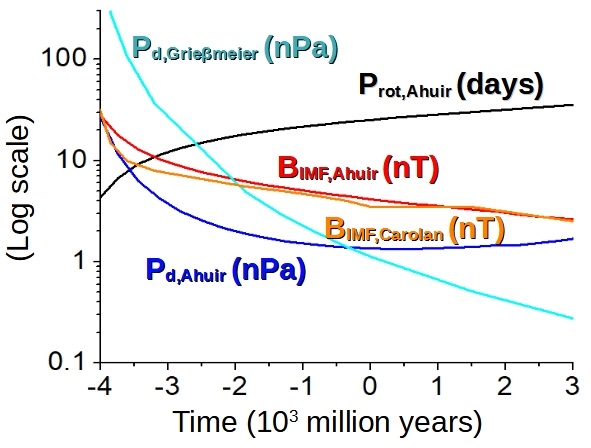}}
\caption{SW dynamic pressure by \citet{Griemeier} (cyan line) and by \citet{Ahuir} (blue line) models. IMF intensity by \citet{Carolan} (orange line) and by \citet{Ahuir} (red line) models. Sun rotation period by \citet{Ahuir} model (black line).}
\label{3}
\end{figure}

\section{Earth habitability during periods with low and high dipolar magnetic field strength}
\label{Dipole}

This section is dedicated to the analysis of the Earth habitability during periods of weak ($\leq 20$ $\mu$T), standard ($\sim 30$ $\mu$T) and high ($\geq 40$ $\mu$T) dipolar magnetic field intensity. The dipole model with $5$ $\mu$T represents low field periods during the Proterozoic eon as well as the Cambrian, Denovian and Carboniferous periods in the Paleozoic era and the Triassic period in the Mesozoic era \citep{Elming,Hawkins,Hawkins2,Thallner,Heunemann}. The dipole model with $15$ $\mu$T is dedicated to study low field periods during the Paleo-Archean and Meso-Archean eras, Proterozoic eon, Jurassic period in the Mesozoic era and the Paleogene period in the Cenozoic era \citep{Morimoto,Chiara,Prevot,Kosterov,Juarez,Rathert}. The dipole model with $30$ $\mu$T illustrates the Meso-Proterozoic and Neo-Archean eras as well as the Neogene and Quaternary periods in the Cretaceous era \citep{Yoshihara,Chiara2,Kaya}. The dipole model with $45$ $\mu$T represents high field periods during the Hadean eon and the Eo-Archean era \citep{Hale,Tarduo4}.

Figure \ref{3} shows the Earth magnetosphere response during an ICME event with $|B|_{IMF} = 50$ nT and $P_{d} = 50$ nPa for a dipole model and different magnetic field intensities. The magnetopause standoff distance is located further away from the Earth surface as the dipole magnetic field intensity increases, from $r_{mp} = 2.14$ if $B_{E} = 5$ $\mu$T to $4.89$ if $B_{E} = 45$ $\mu$T. The simulations show that there is no habitability impact caused by the SW during strong ICME events if the Earth field intensity is at least $5$ $\mu$T. 

\begin{figure}
\centering
\resizebox{\hsize}{!}{\includegraphics[width=\columnwidth]{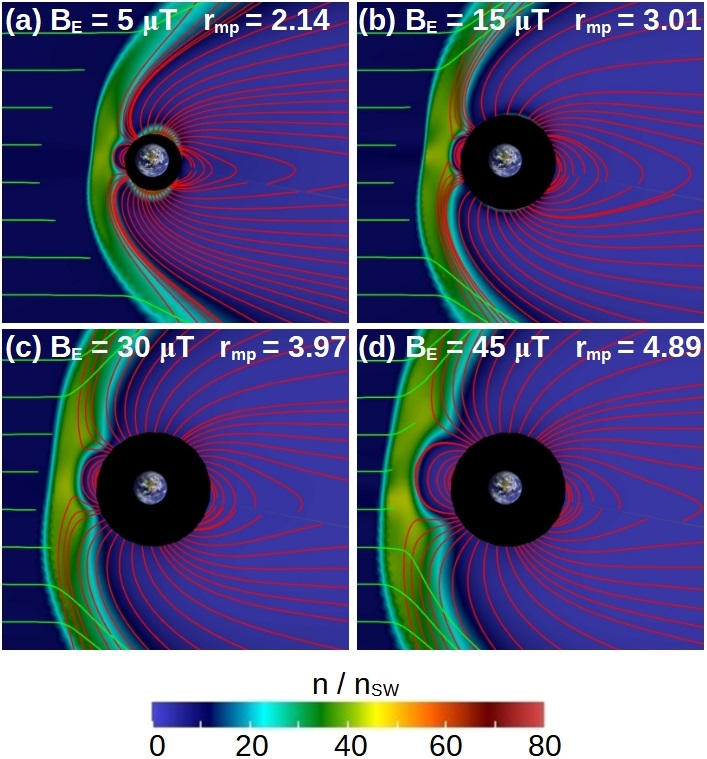}}
\caption{Polar cut of the SW density if the Earth dipole intensity is (a) $5$ $\mu$T, (b) $15$ $\mu$T, (a) $30$ $\mu$T and (a) $45$ $\mu$T. SW dynamic pressure of $50$ nPa, Southward IMF intensity of $50$ nT. The red lines show the Earth magnetic field and the green lines the SW velocity stream lines.}
\label{3}
\end{figure}

Figure \ref{4} shows the magnetopause standoff distance for different space weather conditions (SW dynamic pressure and Southward IMF intensity) and Earth dipole magnetic field intensities. It should be recalled that the inner boundary of the simulation frame is above the Earth surface in all the models analyzed, thus the simulations set cannot include space weather conditions leading to the direct precipitation of the SW towards the Earth surface. The general trend indicates a decrease of $r_{mp}$ as $P_{d}$ and $B_{IMF}$ increase because the magnetosphere compression by the SW is higher and the Earth magnetic shield erosion by the IMF is stronger. The simulations for a dipole with $45$ and $30$ $\mu$T show $r_{mp} > 2.5$ in all the space weather configurations analyzed, that is to say, a dipole with $\geq 30$ $\mu$T can bring an adequate protection during super ICME events ($P_{d} = 250$ nPa and $B_{IMF} = 250$ nT), stronger than the Carrington event \citep{Carrington}. The simulations for a dipole with $15$ and $5$ $\mu$T also show $r_{mp} > 2.25$ and $1.5$ during extreme space weather conditions, respectively.

\begin{figure}
\centering
\resizebox{\hsize}{!}{\includegraphics[width=\columnwidth]{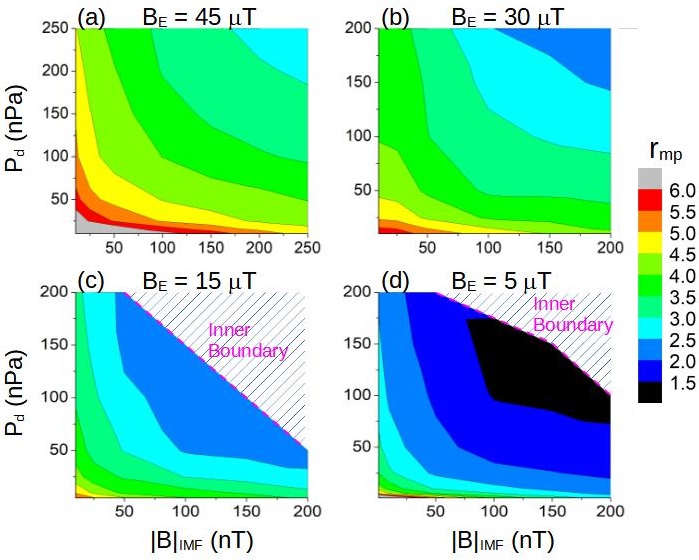}}
\caption{Isocontour of the magnetopause standoff distance for different space weather conditions if the Earth dipole magnetic field intensity is (a) $45$, (b) $30$, (c) $15$, and (d) $5$ $\mu$T.}
\label{4}
\end{figure}

Based on on the parametric analysis performed, the magnetopause standoff distance is fitted with respect to $P_{d}$ and $|B|_{IMF}$ by the surface function $log(r_{mp}) = log(Z) + Mlog(|B|_{IMF}) + Nlog(P_{d})$. The regression provides the exponential dependency of $r_{mp}$ with respect to the  IMF intensity ($M$) and the SW dynamic pressure ($N$) as well as the proportionality constant $Z$. Table \ref{2} lists the coefficients of the regressions. 

\begin{table*}
\centering
\begin{tabular}{c | c c c} \hline
$B_{E}$ ($\mu$T) & $log(Z)$ & $M$ & $N$ \\ \hline
$45$ & $2.84 \pm 0.03$ & $0.141 \pm 0.005$ & $0.181 \pm 0.005$ \\
$30$ & $2.55 \pm 0.05$ & $0.131 \pm 0.009$ & $0.176 \pm 0.009$ \\
$15$ & $2.37 \pm 0.03$ & $0.161 \pm 0.005$ & $0.167 \pm 0.004$ \\
$5$ & $2.38 \pm 0.05$ & $0.21 \pm 0.01$ & $0.204 \pm 0.006$ \\
\end{tabular}
\caption{Regression parameters ($log(r_{mp}) = log(Z) + Mlog(|B|_{IMF}) + Nlog(P_{d})$) in simulations with different SW dynamic pressure, IMF, and dipole intensity. Dipole intensity (first column), $log(Z)$ parameter (second column), $M$ parameter (third column), and $N$ parameter (fourth column). The standard errors of the regression parameters are included.}
\label{2}
\end{table*}

Figure \ref{5} indicates the critical $B_{IMF}$ values for quiet and ICME-like space weather conditions with respect to $P_{d}$ (panels a and b) and time (panels c and d) based on the G-V-C model. The needed combination of $P_{d}$ and $B_{IMF}$ to cause the direct SW deposition requires lower values as the dipole intensity decreases, particularly during ICME events. The large SW dynamic pressure in early stages of the Sun main sequence leads to a sharp decrease of the critical $B_{IMF}$ in the Hadean eon, $0.1$-$340$ nT for a dipole with $5$ $\mu$T, $1.1$-$6.9 \cdot 10^{3}$ nT for a dipole with $15$ $\mu$T, $1.6$-$140 \cdot 10^{3}$ nT for a dipole with $30$ $\mu$T and $7.4$-$390 \cdot 10^{3}$ nT for a dipole with $45$ $\mu$T during quiet space weather conditions. ICME events cause a further decrease of the critical $B_{IMF}$, predicting the inability of a $5$ $\mu$T dipole to protect the Earth, as well as a rather low threshold of $0.01$-$60$ nT for a $15$ $\mu$T dipole, $0.05$-$290$ nT for a $30$ $\mu$T dipole and $0.1$-$1050$ nT for a $45$ $\mu$T dipole. Consequently, the Earth habitability during the Hadean eon was strongly hampered by the space weather conditions, particularly due to the high recurrence of ICME events generated by the young Sun. Quiet space weather conditions during the Archean eon weakens compared to the Hadean eon, thus the critical $B_{IMF}$ increases to $340$-$2.5 \cdot 10^{3}$ nT for a $5$ $\mu$T dipole and above $7 \cdot 10^{3}$ nT for the rest of dipole models. Nevertheless, ICME events still cause a severe decrease of the critical $B_{IMF}$, $4$-$30$ nT for a $5$ $\mu$T dipole, $60$-$500$ nT for a $15$ $\mu$T dipole, $290$-$4.6 \cdot 10^{3}$ nT for a $30$ $\mu$T dipole and $1050$-$15 \cdot 10^{3}$ nT for a $45$ $\mu$T dipole. Thus, the Earth habitability could be impacted in low field periods of the Archean Eon for ICME conditions because the lower bounds of the critical $B_{IMF}$ are rather low in dipole models with $5$ and $15$ $\mu$T. There is a further mitigation of the space weather conditions along the Proterozoic eon leading to a critical $B_{IMF}$ above $10^{4}$ nT for all the dipole intensities tested during quiet space weather conditions. Nevertheless, ICME events can still jeopardize the Earth habitability during periods of low dipolar field because the critical $B_{IMF}$ for the dipole model with $5$ $\mu$T is $200$-$500$ nT, larger than $4000$ nT for the dipole models with $\geq 15$ $\mu$T. Regarding the Phanerozoic and future eons, a dipole with an intensity of $5$ $\mu$T or higher may provide an efficient shielding during quiet and ICME-like space weather conditions because the critical IMF intensity is above $700$ nT, excluding super-flare events. Next, the critical $B_{IMF}$ calculated is compared with the predicted IMF intensity at different geological times. The dipole with $\geq 5$ $\mu$T can protect the Earth for quiet space weather conditions during all the geological times except the Hadean eon, including the young Sun and cyclic modulations, although ICME-like space weather conditions can constrain the Earth habitability. ICME events in the Hadean eon may lead to the direct precipitation of the SW even though the dipole intensity is $45$ $\mu$T. On the other hand, a dipole intensity of $\geq 30$ $\mu$T may provide an efficient shielding from the Archean eon. A $15$ $\mu$T dipole can only provide complete protection from the Meso-Archean era and a $5$ $\mu$T dipole from the Meso-Proterozoic era. Consequently, the Earth surface and atmosphere could be exposed to the SW during ICME events in the Hadean eon and dipolar low field periods in the Paleo-Archean and Paleo-Proterozoic eras.

\begin{figure}
\centering
\resizebox{\hsize}{!}{\includegraphics[width=\columnwidth]{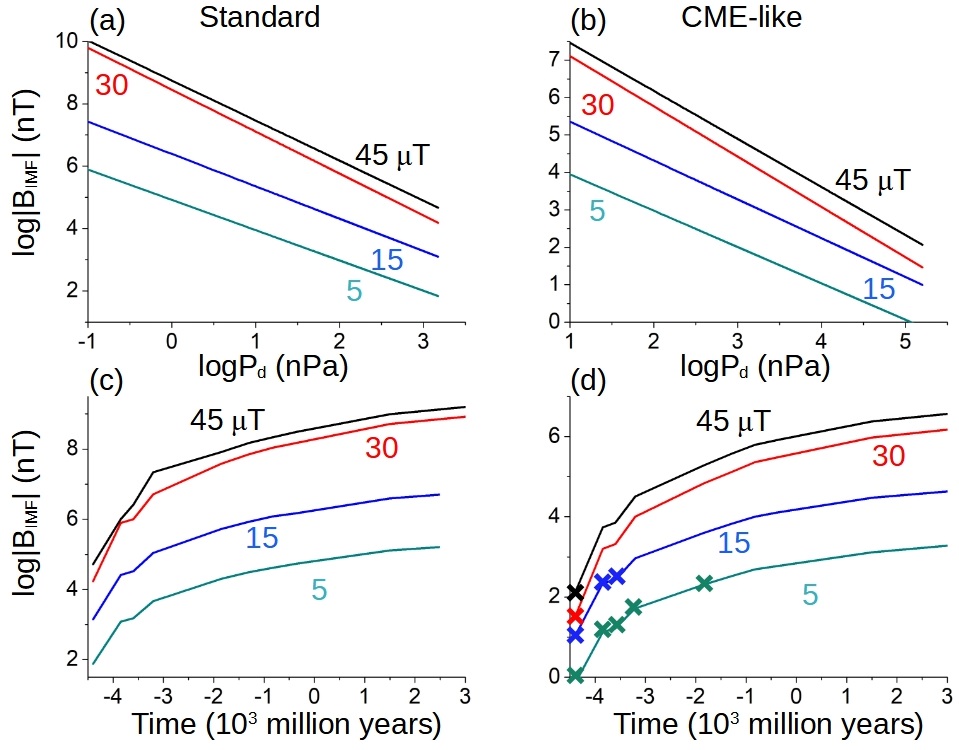}}
\caption{Critical Southwards IMF intensity required in G-V-C model for the direct deposition of the SW towards the Earth surface during quiet space weather conditions with respect to the (a) SW dynamic pressure and (b) geological time. Same study for ICME-like space weather conditions with respect to (b) SW dynamic pressure and (d) geological time. The black line indicates the dipole model with $45$ $\mu$T, the red line for $30$ $\mu$T, the blue line for $15$ $\mu$T and the cyan line for $5$ $\mu$T. The bold crosses indicate the geological times the SW may jeopardize the Earth habitability for each magnetic field configuration.}
\label{5}
\end{figure}

Figure \ref{6} shows the same analysis although performed using the space weather conditions predicted by the Ahuir model. Quiet space weather conditions are not included in the discussion because there is no strong impact on the Earth habitability. The analysis shows some differences compared to G-V-C model results. In particular, the analysis based on Ahuir model indicates a $45$ $\mu$T dipole can protect the Earth during the Hadean eon, unprotected based on G-V-C model. Likewise, $15$ and $5$ $\mu$T dipoles provides an efficient shielding from the Eo-Archean and Paleo-Proterozoic eras, respectively, although G-V-C model delay the magnetosphere protection to the Meso-Archean and Meso-Proterozoic eras. The deviation between models results is caused by a lower SW dynamic pressure estimation in the Ahuir model before the Neo-Proterozoic.

\begin{figure}
\centering
\resizebox{\hsize}{!}{\includegraphics[width=\columnwidth]{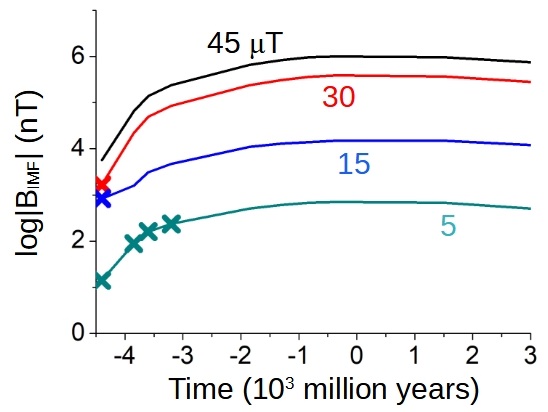}}
\caption{Critical Southwards IMF intensity required in Ahuir model for the direct deposition of the SW towards the Earth surface during ICME-like space weather conditions for a dipolar magnetic field configuration. black line $45$ $\mu$T, red line $30$ $\mu$T,  blue line $15$ $\mu$T and cyan line $5$ $\mu$T. The bold crosses indicate the geological times the SW may jeopardize the Earth habitability for each magnetic field configuration.}
\label{6}
\end{figure}

In summary, habitability constraints caused by the SW during the Hadean eon did not affect the life evolution because the Earth was inhospitable to any organisms, showing acid boiling oceans and an atmosphere rich in carbon dioxide as well as the continuous bombardment of asteroids and comets \citep{Kasting3,Morse,Dodd}. On the other hand, the erosion caused by the SW on the primary Earth atmosphere could enhance the water molecules escape as well as the fast Hydrogen and Helium exhaust by increasing the leak ratios \citep{Shaw,Kasting4,Thomassot}. SW sterilizing effect during the Paleo-Archean may have an effect on the emerging life such as microbial mats and bacteria \citep{Lepot} as well as on the atmosphere evolution \citep{Catling}. Same discussion can be done for the Meso-Proterozoic era, particularly about the effect of the SW on the outbreak of complex life forms as multicellular organisms \citep{Butterfield} and the atmosphere early oxygenation \citep{Parnell}.

\section{Earth habitability during periods with multipolar magnetic field}
\label{Multipolar}

This section is dedicated to study the Earth habitability during periods of weak multipolar fields linked to geomagnetic reversals \citep{McElhinny2,Bogue,Gubbins}. Polarity inversions lead to a weakening of the Earth magnetic shield thus weaker space weather conditions are required for the SW direct precipitation towards the surface \citep{Glatzmaiers}. For example, paleomagnetic data revealed a large quadrupolar component of the Earth magnetic field along the Proterozoic age as well as the Devonian and Cretaceous periods \citep{Lhuillier,Shcherbakova,Guyodo}. There is a limited amount of information about the magnetic field topology during geomagnetic reversals \citep{Valet2,Valet3}, thus the effect of the space weather on the habitability is modeled using four different configurations, multipolar fields with quadrupole to dipole coefficient ratios of $0.2$ and $0.5$ for a field intensity of $5$ and $15$ $\mu$T at the equatorial region of the Earth surface. Consequently, the simulations results can be directly compared with the dipolar models with $5$ and $15$ $\mu$T.

\begin{figure}
\centering
\includegraphics[width=8cm]{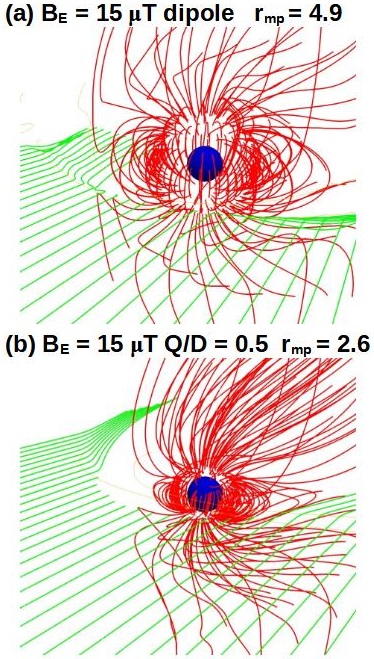}
\caption{3D view of the Earth magnetosphere for (a) dipole and (b) multipolar $Q/D=0.5$ models with a magnetic field intensity of $15$ $\mu$T. Earth magnetic field lines (red lines) and SW velocity stream lines (green lines).}
\label{7}
\end{figure}

Figure \ref{7} shows the deformation of the Earth magnetosphere caused by a common ICME ($P_{d} = 10$ nPa and $B_{IMF} = 10$ nT) during a dipolar low field period, panel a, and a multipolar low field period with $Q/D=0.5$, panel b, if the field intensity is $15$ $\mu$T. The magnetic field lines (red lines) show clear differences in the magnetosphere topology and the SW deflection by the magnetic shield (green lines indicate the SW velocity stream lines). Consequently, the magnetopause standoff distance is different, $r_{mp} = 4.9$ in the dipole model and $2.9$ in the multipolar model, pointing out a rather large decrease of $r_{mp}$ in the multipolar case compared to the dipole model. That means a multipolar Earth magnetic field is strongly constrained by the space weather conditions and the Earth habitability easily hampered.

\begin{figure}
\centering
\resizebox{\hsize}{!}{\includegraphics[width=\columnwidth]{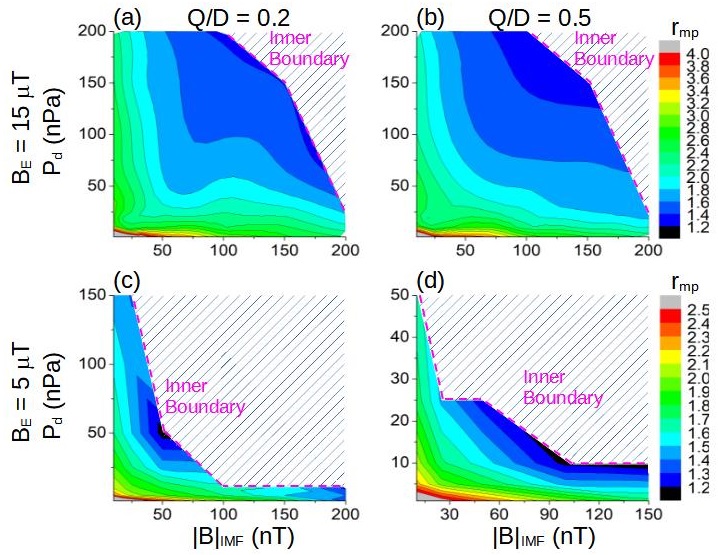}}
\caption{Isocontour of the magnetopause standoff distance for different space weather conditions if (a) $Q/D = 0.2$ and $B_{E}=15$ $\mu$T, (b) $Q/D = 0.5$ and $B_{E}=15$ $\mu$T, (c) $Q/D = 0.2$ and $B_{E}=5$ $\mu$T, (d) $Q/D = 0.5$ and $B_{E}=5$ $\mu$T.}
\label{8}
\end{figure}

Figure \ref{8} shows the magnetopause standoff distance for different space weather conditions, Earth magnetic field quadrupole to dipole coefficient ratios and intensities. If the panels a and b are compared with the figure \ref{4}c, there is a decrease of $r_{mp}$ as the $Q/D$ ratio increases for the same space weather conditions, particularly as the ICME events are more severe. The same discussion can be done comparing panels c and d with the figure \ref{4}d. In addition, the multipolar model with $15$ $\mu$T $Q/D = 0.2$ shows $r_{mp}$ values $15 \%$ larger in averaged compared to the dipolar case with $5$ $\mu$T for the same space weather conditions, reduced to $11 \%$ for the multipolar model with $15$ $\mu$T $Q/D = 0.5$. Consequently, for the same Earth magnetic field intensity, configurations with a larger $Q/D$ ratio lead to a lower magnetopause stand off distance. Indeed, ICME-like space weather conditions leads to $r_{mp} < 2.0$, thus the Earth habitability is more hampered in the multipolar models with respect to the dipole model. 

Next, $r_{mp}$ values calculated in the parametric analysis are fitted using the same surface function of section \ref{Dipole}. Table \ref{3} indicates the coefficients of the regressions.

\begin{table*}
\centering
\begin{tabular}{c c | c c c} \hline
$Q/D$ & $B_{E}$ ($\mu$T) & $log(Z)$ & $M$ & $N$ \\ \hline
$0.2$ & $15$ & $1.95 \pm 0.06$ & $0.175 \pm 0.013$ & $0.144 \pm 0.008$ \\
$0.5$ & $15$ & $1.88 \pm 0.04$ & $0.164 \pm 0.009$ & $0.151 \pm 0.005$ \\
$0.2$ & $5$ & $1.52 \pm 0.07$ & $0.18 \pm 0.02$ & $0.13 \pm 0.01$ \\
$0.5$ & $5$ & $1.58 \pm 0.06$ & $0.201 \pm 0.013$ & $0.172 \pm 0.011$ \\
\end{tabular}
\caption{Regression parameters ($log(r_{mp}) = log(Z) + Mlog(|B|_{IMF}) + Nlog(P_{d})$) in simulations with different SW dynamic pressure, IMF intensity, $Q/D$ ratio, and Earth magnetic field intensity. $Q/D$ ratio (first column), Earth magnetic field intensity (second column), $log(Z)$ parameter (third column), $M$ parameter (fourth column), and $N$ parameter (fifth column). The standard errors of the regression parameters are included.}
\label{3}
\end{table*}

Again, the critical $B_{IMF}$ required for the direct SW deposition is calculated based on the SW velocity and density estimation at the Earth orbit along the Sun's evolution on the main sequence by G-V-C model. Figure \ref{8} shows the critical $B_{IMF}$ values for quiet and ICME-like space weather conditions with respect to $P_{d}$ and time. The critical $B_{IMF}$ in the Hadean eon during quiet space weather conditions for the models with $5$ $\mu$T $Q/D = 0.2$ and $0.5$ is $0.2-80$ and $0.1$-$20$ nT, respectively. On the other hand, the Earth is completely exposed to the SW during ICME events. The models with $15$ $\mu$T $Q/D = 0.2$ and $0.5$ show an increase of the critical $B_{IMF}$ by one order of magnitude compared to $5$ $\mu$T cases for quiet space weather conditions, $1-520$ nT, although the critical $B_{IMF}$ is rather lower during ICMEs, $\leq 0.1$ nT. Consequently, multipolar low fields cannot protect the Earth during the Hadean eon. A more tempered space weather in the Archean eon lead to an increase of the critical $B_{IMF}$ to $20-120$ nT for the model with $5$ $\mu$T $Q/D = 0.5$ and to $80-345$ nT for the $5$ $\mu$T $Q/D = 0.2$ case during quiet conditions. The critical $B_{IMF}$ is below $50$ nT during ICME events so multipolar $5$ $\mu$T models cannot provide an efficient magnetic shield. Same discussion can be done for the $15$ $\mu$T models, enough to protect the Earth during quiet space weather condition, critical $B_{IMF}$ ranges between $1$-$3425$ nT, although decreasing to values $\leq 1$-$50$ nT during ICME events. Thus, multipolar low fields may not be able to protect the Earth during the Archeon eon neither. Proterozoic and Phanerozoic eons show critical $B_{IMF}$ values above $660$ nT for both $5$ $\mu$T models during quiet space weather conditions. The lower-upper bounds during ICME-events is $15$-$46$ nT in the model with $Q/D = 0.5$ and $53$-$154$ nT if $Q/D = 0.2$, thus multipolar low fields with $5$ $\mu$T may not protect the Earth during strong ICME events even in the present days. On the other hand, $15$ $\mu$T models may provide an efficient shield during quiet and ICME events. Regarding Posterum eons, multipolar fields with $15$ $\mu$T $Q/D \leq 0.5$ and $5$ $\mu$T $Q/D \leq 0.2$ can protect the Earth, critical $B_{IMF} \approx 270$ nT, although $\leq 5$ $\mu$T $Q/D \geq 0.5$ magnetic fields may not provide an efficient shielding for Carrington-like events with a critical $B_{IMF} \approx 90$ nT. Now, the critical $B_{IMF}$ is compared with the predicted $B_{IMF}$ values along the Sun's evolution on the main sequence. Quiet space weather conditions can only hamper the Earth habitability during the Hadean eon if the mutipolar magnetic field intensity is $\leq 15$ $\mu$T. ICME events could constrain the Earth habitability in the Archean eons as well as in the Paleo-Proterozoico era during multipolar field periods with $B_{E} \leq 15$ $\mu$T and $Q/D \geq 0.2$. From the Meso-Proterozoic era multipolar fields with $15$ $\mu$T and $Q/D \leq 0.5$ provide an efficient magnetic shield, just like a multipolar field with $5$ $\mu$T and $Q/D \leq 0.2$ in Posterum eons. Only the multipolar fields with $5$ $\mu$T and $Q/D \geq 0.2$ are unable to protect the Earth in the Paleozoic era and Posterum eons.

\begin{figure}
\centering
\resizebox{\hsize}{!}{\includegraphics[width=\columnwidth]{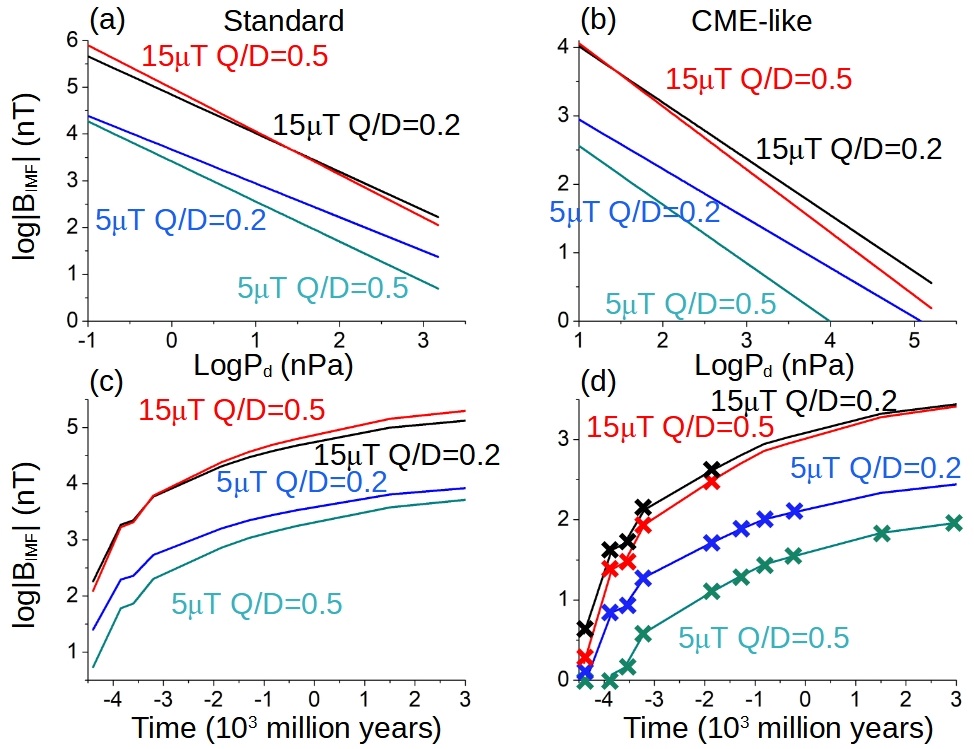}}
\caption{Critical Southwards IMF intensity required in G-V-C model for the direct deposition of the SW towards the Earth surface during quiet space weather conditions with respect to the (a) SW dynamic pressure and (c) geological time. Same study for ICME-like space weather conditions with respect to the (b) SW dynamic pressure and (d) geological time. The black line indicates the model with $Q/D = 0.2$ and $15$ $\mu$T, the red line for $Q/D = 0.5$ and $15$ $\mu$T, the blue line for $Q/D = 0.2$ and $5$ $\mu$T and the cyan line for $Q/D = 0.5$ and $5$ $\mu$T. The bold crosses indicate the geological times the SW may jeopardize the Earth habitability for each magnetic field configuration.}
\label{9}
\end{figure}

\begin{figure}
\centering
\resizebox{\hsize}{!}{\includegraphics[width=\columnwidth]{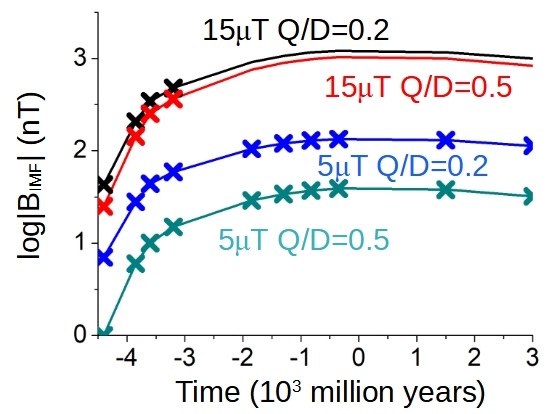}}
\caption{Critical Southwards IMF intensity required in Ahuir model for the direct deposition of the SW towards the Earth surface during ICME-like space weather conditions for a multipolar Earth magnetic field configuration: black line $Q/D = 0.2$ and $15$ $\mu$T, red line $Q/D = 0.5$ and $15$ $\mu$T, blue line $Q/D = 0.2$ and $5$ $\mu$T and cyan line $Q/D = 0.5$ and $5$ $\mu$T. The bold crosses indicate the geological times the SW may jeopardize the Earth habitability for each magnetic field configuration.}
\label{10}
\end{figure}

Figure \ref{10} shows the analysis results based on Ahuir space weather predictions. A multipolar field with $15$ $\mu$T and $Q/D \leq 0.5$ can protect the Earth from the Paleo-Proterozoic era based on Ahuir model (from the Meso-Proterozoic era based on G-V-C models) because the predicted SW dynamic pressure is lower. On the other hand, a multipolar field with $5$ $\mu$T and $Q/D \geq 0.2$ cannot shield the Earth during any geological time using Ahuir model, unlike the G-V-C models that show an efficient protection from the Paleozoic for the configuration with $Q/D \leq 0.2$. Such difference is explained by an increment of the SW dynamic pressure from the Paleozoic eon in Ahuir model, leading to a larger SW dynamic pressure with respect to the G-V-C models from the Neo-Proterozoic era.

The analysis of the multipolar low field periods confirms the discussion initiated in section \ref{Dipole} about the Earth vulnerability during geomagnetic reversals. Multipolar low field periods may lead to an enhanced atmosphere erosion during the Hadean and Archean eons as well as the Paleo-Proterozoic era. Also, multipolar low field periods with $B_{E} \leq 5$ $\mu$T during the Proterozoic eon could affect the incipient multicellular organisms. It should be noted that multipolar low field periods with $B_{E} \leq 5$ $\mu$T and $Q/D \geq 0.5$ are unable to protect the Earth in the Phanerozoic eon, that is to say, the Earth habitability can be jeopardized during geomagnetic reversals in the present time. Indeed, several studies link mass-extinction in the Phanerozoic eon with geomagnetic reversal periods \citep{Kennett,Plotnick,Worm,Isozaki,Wei,Isozaki2}, particularly for the Late Devonian \citep{McGhee,Hawkins5,Boon} ($372$ million years ago), Permian–Triassic ($251.9$ million years ago) \citep{Benton,Heunemann,Poreda,Onoue} and Triassic–Jurassic ($201.4$ million years ago) \citep{Zijl,Wei,Schoepfer} extinction events.

\section{Earth habitability in Posterum eons: decaying dipole}
\label{Posterum}

The last section of the study is dedicated to analyze the effect of the SW on the Earth habitability in Posterum eons, once the geodynamo begins to fade away as well as the Earth magnetic field. Numerical models predict a decay of the Earth magnetic field as the Earth core cools down and expands leading to the eventual termination of the geodynamo, a process that occurred several thousand million years ago in Mars \citep{Roberts,Roberts2,Jones,Smirnov}. Geodynamo simulations predict large non dipolar components of the Earth magnetic field during the decay phase \citep{Roberts}, thus the analysis is connected with the multipolar models studied in the previous section that already identified a vulnerable Earth habitability during ICME-like events if $B_{E} \leq 5$ $\mu$T and $Q/D \geq 0.5$. Consequently, a further decrease of the Earth magnetic field intensity or an increase of the $Q/D$ ratio along Posterum eons could cause an inefficient magnetic shielding not only during ICME event, also for quiet space weather conditions. Such hypothetical configurations are explored performing a new set of simulations assuming Earth dipolar fields with $1$ and $0.1$ $\mu$T, providing the upper bound of the Earth habitability with respect to multipolar models for the same field intensity. In addition, the analysis is extended to include different IMF orientations, particularly the Sun-Earth or radial IMF (parallel to the SW velocity vector), Northward and Southward IMF (perpendicular to the SW velocity vector at the XZ plane), and ecliptic counterclockwise IMF.

\begin{figure}
\centering
\includegraphics[width=8cm]{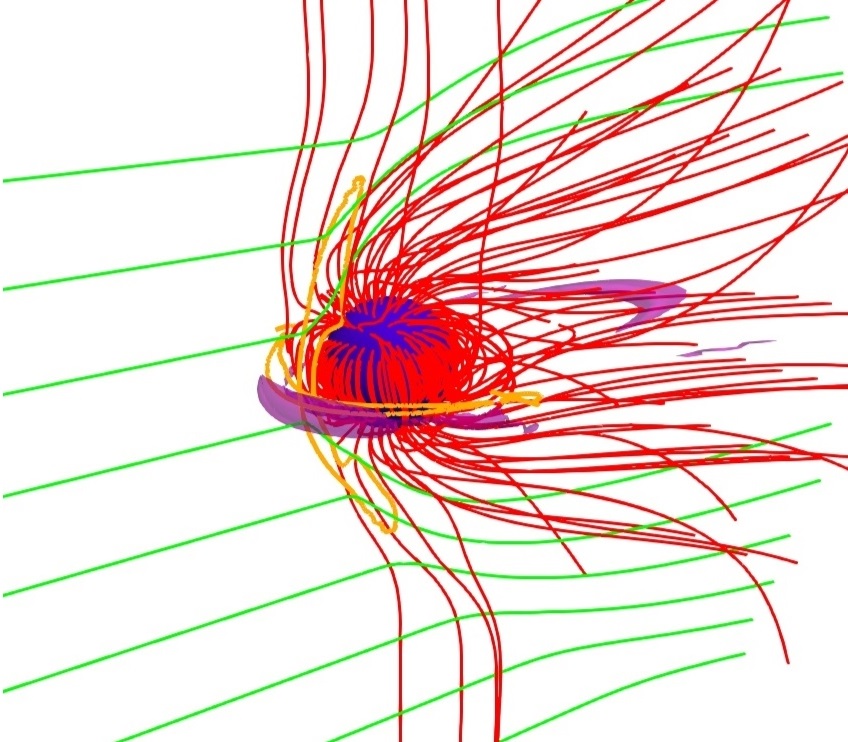}
\caption{3D view of the Earth magnetosphere topology for a dipole $0.1$ $\mu$T during quiet space weather conditions (Southward $B_{IMF} = 1$ nT and $P_{d} = 1$ nPa). Earth magnetic field lines (red lines), SW velocity stream lines (green lines), isoline of the plasma density for $n/n_{sw}=30$ in the XY and XZ planes (bold orange lines), and reconnection region between the IMF and the Earth magnetic field (pink iso-volume.)}
\label{11}
\end{figure}

Figure \ref{11} shows the Earth magnetosphere during quiet space weather conditions ($P_{d} = 1$ nPa and Southward $B_{IMF} = 1$ nT) if the Earth dipole intensity is $0.1$ $\mu$T. The magnetopause standoff distance is $1.44$, that is to say, the Earth magnetic field (red lines) can barely deflect the SW (green lines) thus the bow show (orange lines) and the reconnection region between IMF and Earth magnetic field (pink iso-volume) are located close to the Earth surface. Consequently, the Earth surface and atmosphere could be exposed to the SW as soon as there is a small increase of the SW density / velocity or the IMF intensity.

\begin{figure}
\centering
\resizebox{\hsize}{!}{\includegraphics[width=\columnwidth]{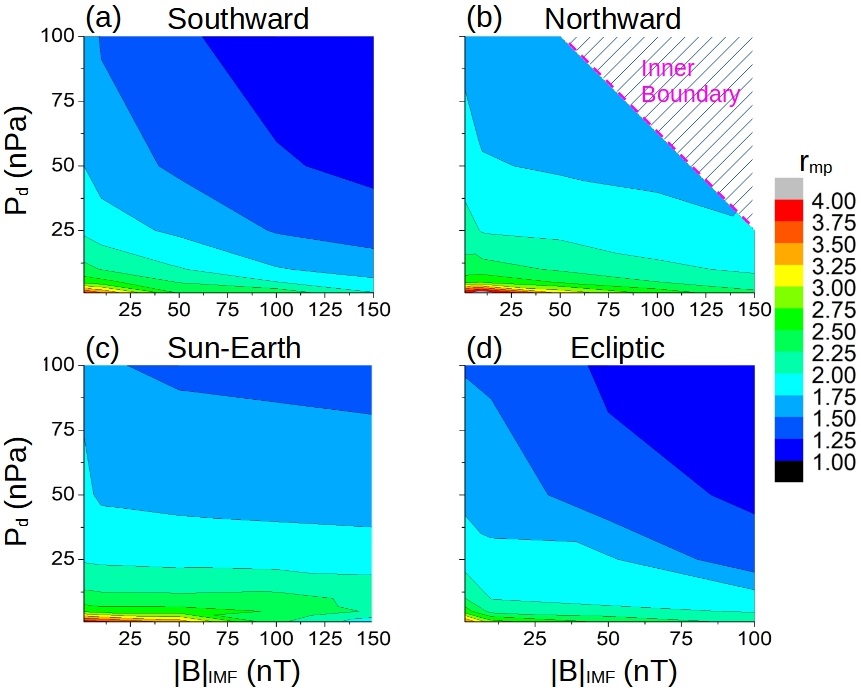}}
\caption{Isocontour of the magnetopause standoff distance for different space weather conditions if the Earth dipole intensity is $1$ $\mu$T and the IMF orientation is (a) Southward (b) Northward, (c) Sun-Earth, and (d) Ecliptic.}
\label{12}
\end{figure}

\begin{figure}
\centering
\resizebox{\hsize}{!}{\includegraphics[width=\columnwidth]{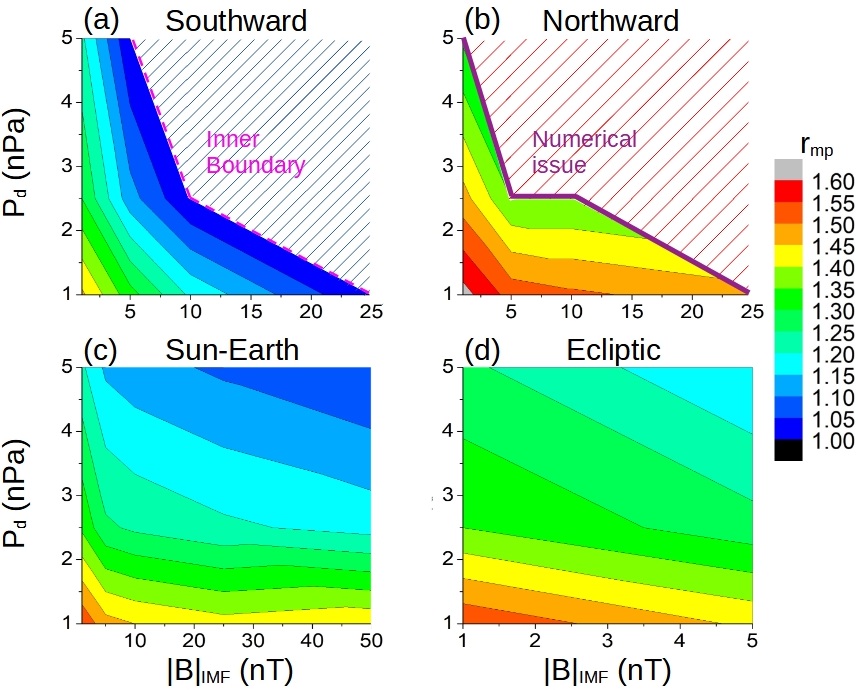}}
\caption{Isocontour of the magnetopause standoff distance for different space weather conditions if the Earth dipole intensity is $0.1$ $\mu$T and the IMF orientation is (a) Southward (b) Northward, (c) Sun-Earth, and (d) Ecliptic.}
\label{13}
\end{figure}

Figure \ref{12} and \ref{13} show the magnetopause standoff distance for different space weather conditions and dipole models with $1$ and $0.1$ $\mu$T, respectively. The erosion caused by a Southward IMF in the Earth magnetic field leads to the configurations with the lowest $r_{mp}$. On the other hand, a Northward IMF enhances the Earth magnetic field in the equatorial region leading to the configuration with the largest $r_{mp}$. The Sun-Earth IMF erodes the Earth magnetic field in the Southward hemisphere and the ecliptic IMF induces a magnetosphere East-West tilt, leading to configuration showing intermediate $r_{mp}$ values compared to the Southward and Northward IMF cases. Dipole model simulations with $1$ $\mu$T show $r_{mp} > 1$ for space weather configurations with $P_{d} \leq 100$ nPa and $B_{IMF} \leq 100$ nT for all the IMF orientations. On the other hand, if the dipole intensity is $0.1$ $\mu$T, configurations leading to $r_{mp} > 1$ requires $P_{d}$ and $B_{IMF}$ values not much higher than quiet space weather conditions.

The $r_{mp}$ values calculated for the dipole model with $1$ and $0.1$ $\mu$T are fitted using the same surface function of section \ref{Dipole}. The regression results are shown in the table \ref{4}.

\begin{table*}
\centering
\begin{tabular}{c c | c c c} \hline
$B_{E}$ ($\mu$T) & IMF & $log(Z)$ & $M$ & $N$ \\ \hline
$1$  & Southward &  $1.55 \pm 0.04$ & $0.119 \pm 0.0008$ & $0.194 \pm 0.008$ \\
$1$ & Northward & $1.32 \pm 0.05$ & $0.054 \pm 0.009$ & $0.150 \pm 0.009$ \\
$1$ & Earth-Sun & $1.31 \pm 0.04$ & $0.024 \pm 0.007$ & $0.180 \pm 0.007$ \\
$1$ & Ecliptic & $1.32 \pm 0.09$ & $0.08 \pm 0.02$ & $0.17 \pm 0.02$ \\
$0.1$ & Southward &  $0.36 \pm 0.03$ & $0.088 \pm 0.013$ & $0.10 \pm 0.02$ \\
$0.1$ & Northward & $0.485 \pm 0.012$ & $0.032 \pm 0.005$ & $0.13 \pm 0.01$ \\
$0.1$ & Earth-Sun & $0.441 \pm 0.008$ & $0.045 \pm 0.003$ & $0.166 \pm 0.005$ \\
$0.1$ & Ecliptic & $0.433 \pm 0.008$ & $0.045 \pm 0.005$ & $0.133 \pm 0.007$ \\
\end{tabular}
\caption{Regression parameters ($log(r_{mp}) = log(Z) + Mlog(|B|_{IMF}) + Nlog(P_{d})$) in simulations with different SW dynamic pressure, IMF intensity and orientation as well as Earth magnetic field intensity. Earth magnetic field intensity (first column), IMF orientation (second column), $log(Z)$ parameter (third column), $M$ parameter (fourth column), and $N$ parameter (fifth column). The standard errors of the regression parameters are included.}
\label{4}
\end{table*}

\begin{figure}
\centering
\resizebox{\hsize}{!}{\includegraphics[width=6cm]{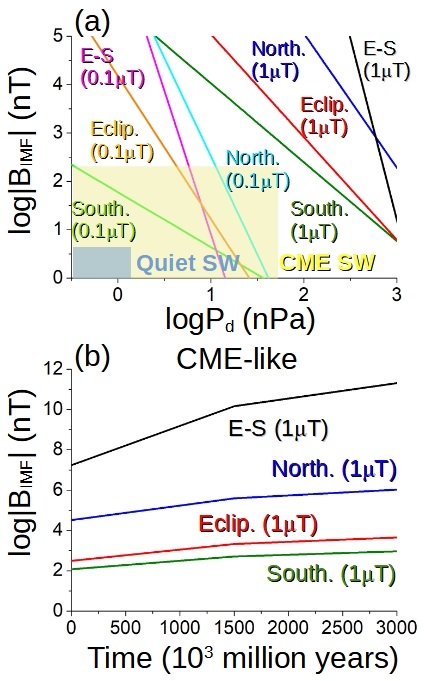}}
\caption{Critical IMF intensity required in the G-V-C model for the direct deposition of the SW towards the Earth surface during ICME space weather conditions with respect to the (a) SW dynamic pressure and (b) geological time (Posterum eons). The green (light green) line indicates the model with $B_{E}=1$ ($0.1$) $\mu$T and Southward IMF, the red (orange) line the model with $B_{E}=1$ ($0.1$) $\mu$T and Ecliptic IMF, the blue (cyan) line the model with $B_{E}=1$ ($0.1$) $\mu$T and Northward IMF, the light black (pink) line indicates the model with $B_{E}=0.1$ ($0.1$) $\mu$T and Earth-Sun IMF. The bold crosses indicate the geological times the SW may jeopardize the Earth habitability for each magnetic field configuration.}
\label{14}
\end{figure}

Figure \ref{14} shows the critical $B_{IMF}$ values for ICME-like space weather conditions with respect to $P_{d}$ and time in the Posterum eons based on the G-V-C model. Quiet space weather conditions are not included in the analysis because there is no impact on the Earth habitability (parametric region highlighted by the light blue rectangle in panel a). The critical $B_{d}$ calculated for a dipole with $0.1$ $\mu$T during ICME events (yellow rectangle in panel a) is smaller than $1$ nT for all the IMF orientation tested. On the other hand, the critical $B_{d}$ for a dipole with $1$ $\mu$T is $750$ nT for a Southward IMF, and $> 10^{4}$ nT for the other IMF orientations. Now, the critical $B_{IMF}$ calculated for different IMF orientations is compared with the predicted $B_{IMF}$ values. If the dipole intensity is $\leq 1$ $\mu$T, a Southward IMF orientation can impact the Earth habitability in present and Posterum eons although a dipole with $ 1$ $\mu$T is strong enough to shield the Earth for all the other IMF orientations analyzed. On the other hand, a dipole with $0.1$ $\mu$T cannot protect the Earth for any of the IMF orientations tested.

\begin{figure}
\centering
\resizebox{\hsize}{!}{\includegraphics[width=6cm]{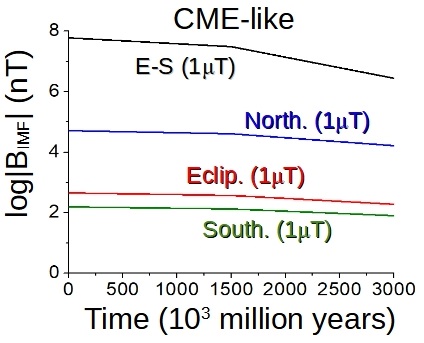}}
\caption{Critical IMF intensity required in the Ahuir model for the direct deposition of the SW towards the Earth surface during ICME space weather conditions with respect to the geological time (Posterum eons). The green (light green) line indicates the model with $B_{E}=1$ ($0.1$) $\mu$T and Southward IMF, the red (orange) line the model with $B_{E}=1$ ($0.1$) $\mu$T and Ecliptic IMF, the blue (cyan) line the model with $B_{E}=1$ ($0.1$) $\mu$T and Northward IMF, the light black (pink) line indicates the model with $B_{E}=0.1$ ($0.1$) $\mu$T and Earth-Sun IMF.}
\label{15}
\end{figure}

Figure \ref{15} shows the analysis results based on Ahuir space weather predictions. The results are very similar compared to the G-V-C model, no impact on the Earth habitability for all the IMF orientations during quiet space weather conditions if the dipole intensity is $\geq 0.1$ $\mu$T. On the other hand, the Earth habitability can be jeopardized during ICME events if the dipole intensity is $\leq 0.1$ $\mu$T for all the IMF orientations analyzed. Only one discrepancy is observed with the G-V-C model results, because the analysis based on the Ahuir models shows ICME-like space weather conditions do not impact the Earth for any of the IMF analyzed if the dipole intensity is $\geq 1$ $\mu$T.

In summary, the analysis shows the Earth habitability could be hampered in Posterum eons during ICME-like space weather conditions if the dipole intensity decreases below $1$ $\mu$T, particularly if the IMF is Southward. This result complements the identification of a direct SW precipitation during geomagnetic reversals obtained in the previous section for multipolar magnetic fields with $\leq 5$ $\mu$T and $Q/S \geq 0.5$. In addition, if the dipole intensity is $\leq 0.1$ $\mu$T, the Earth is barely protected during quiet space weather conditions.

\section{Ionosphere expansion: consequence for the Earth habitability}
\label{Ionosphere}

EUV and X-ray fluxes in the Hadean and Archean eons were much larger compared to the present because the Sun was a more active star \citep{Ribas,Tu}. Decay laws predict a decrease of the X-ray fluxes by a factor of $1000-200$, EUV around hundreds and UV by a factor of $10$ along the Sun main sequence \citep{Guinan,Ribas}. For example, observations by \citet{Guinan} show X-ray fluxes decrease from $3$ to $0.02$ erg s$^{-1}$ cm$^{-2}$ \textup{~\AA}$^{-1}$ comparing star with ages of $130$ and $750$ Mya, further reduced to $10^{-4}$ erg s$^{-1}$ cm$^{-2}$ \textup{~\AA}$^{-1}$ if the star age is $1600$ Mya. Likewise, EUV fluxes decay from $0.3$ to $0.04$ erg s$^{-1}$ cm$^{-2}$ \textup{~\AA}$^{-1}$ between stars with ages of $130$ and $750$ Mya, around $0.004$ erg s$^{-1}$ cm$^{-2}$ \textup{~\AA}$^{-1}$ if the star age is $1600$ Mya. EUV and X-ray radiation are absorbed in the atmosphere upper layers leading to an enhancement of the gas ionization and heating \citep{Tian,Tian2,Johnstone3}. Consequently, the upper atmosphere and ionosphere expanded beyond several Earth radius \citep{Lammer3}, effect also observed in the ionosphere of Mars and Venus \citep{Dubinin,Han}.

The expansion of the upper atmosphere / ionosphere had several consequences, for example an increase of the ions \citep{Lichtenegger,Kislyakova} and neutrals \citep{Tian,Tian2,Johnstone3} loss rates as well as affecting the bow shock and magnetopause standoff distances. That means, the analysis of the Earth habitability with respect to the SW direct deposition in the Hadean and Archean eons requires adding the effect of the ionosphere expansion to improve the model accuracy.

\begin{figure}
\centering
\resizebox{\hsize}{!}{\includegraphics[width=6cm]{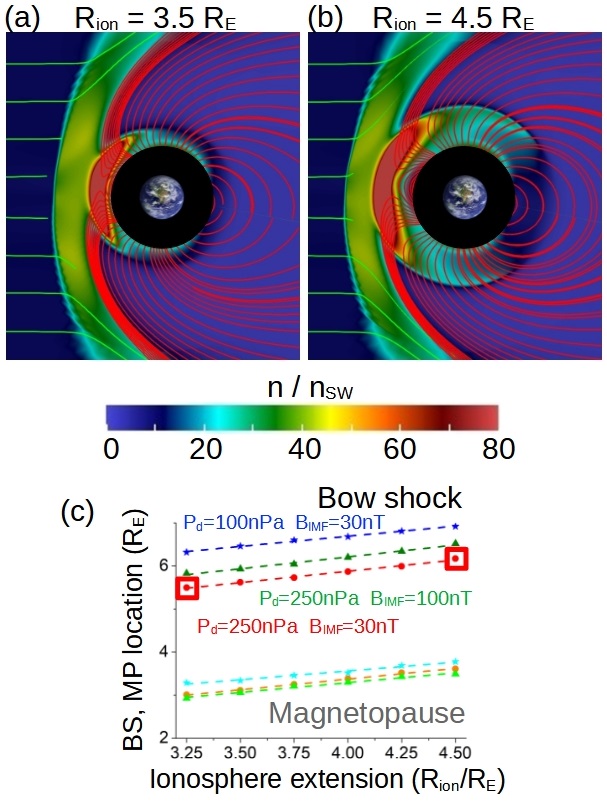}}
\caption{Polar cut of the SW density if the ionosphere extension is (a) $6400$ km and (b) $12740$ km. SW dynamic pressure of $250$ nPa, Southward IMF intensity of $30$ nT. The red lines show the Earth magnetic field and the green lines the SW velocity stream lines. (c) Bow shock ($r_{bs}$) and magnetopause ($r_{mp}$) standoff distances for different ionosphere extensions ($R_{ion}$) and space weather conditions: $P_{d} = 250$ nPa and $B_{IMF} = 30$ nT (red BS and orange MP), $P_{d} = 100$ nPa and $B_{IMF} = 30$ nT (blue BS and cyan MP) and $P_{d} = 250$ nPa and $B_{IMF} = 100$ nT (green BS and light green MP). Red squares indicate the cases plotted in panels a and b.}
\label{A1}
\end{figure}

The ionosphere expansion is introduced in the simulations by modifying the ionosphere model (please see the appendix for further information). In this first attempt, the direct modeling of EUV and X-ray forcing is not included in a self consistent manner, a task that will be performed in a future work. In addition, the analysis implicitly assumes an averaged effect of the different radiation wavelengths on the ionosphere expansion (Paul, Strugarek \& Vaidya 2023 submitted to JGR; Gillet, Garcia-Muñoz \& Strugarek 2023 submitted to A\&A). The extended model can evaluate the effect of the ionosphere expansion on the bow shock and magnetopause standoff distance. A set of simulations are performed dedicated to analyze an ionosphere expanded from $R = 2.5 R_{E}$ to $R = 3.0 R_{E} + i \cdot 0.25 R_{E}$ with $i=[1,6]$, that is to say, ionosphere extensions from $12740$ to $4780$ km representing the Earth ionosphere evolution from the Hadean to the Archean eon as the EUV and X-ray fluxes decays as the Sun evolves along the main sequence \citep{Ribas}. The ionosphere extension is defined as $R_{ion}$. Figure \ref{A1}, panels a and b, show the bow shock and Earth magnetic field lines in the polar plane in simulations with an ionosphere expanded from $2.5$ to $ 3.5 R_{E}$ and from $2.5$ to $4.5 R_{E}$, respectively, if $P_{d} = 250$ nPa and $B_{IMF} = 30$ nT. The bow shock and magnetopause standoff distances increase as the ionosphere expands. Panel c indicates the bow shock and magnetopause standoff distances calculated for different ionosphere extensions and space weather conditions. The analysis indicates the standoff distance increases as the ionosphere expands. The linear fit $r_{MP,BS} = A + B R_{ion}$ is calculated using the bow shock and magnetopause standoff distance with respect to the ionosphere extension. The regression parameters are in table \ref{5}. The ratio between standoff distances and $R_{ion}$ is similar for all the cases tested, and averaged ratio of $0.52$ for the bow shock and $0.46$ for the magnetopause. It should be noted that the simulation results may depend on the IMF orientation, thus a detail characterization requires a more extended analysis out of the scope of the present study. Nevertheless, the simulations already indicate the standoff ratio depends on the bow shock and Earth magnetic field compression induced by the SW dynamic pressure as well as the IMF intensity. The effect of the magnetosphere compression is observed comparing configurations with the same $B_{IMF}$ although different $P_{d}$: the case with $250$ nPa (red-orange) shows lower standoff distances compared to the configuration with $100$ nPa (blue-cyan). Also, the standoff ratio is larger in the case with $250$ nPa because bow shock and magnetopause are closer to the ionosphere and the effect of the ionosphere expansion is larger. The effect of the IMF intensity is analyzed comparing configurations with the same $P_{d}$ and different $B_{IMF}$. The case with $100$ nT shows a lower magnetopause standoff distance with respect to the case with $30$ nT, caused by a stronger erosion of the Earth magnetic field in the equatorial region. However, the bow shock standoff distance is larger in the case with $100$ nT because the magnetic field intensity in the bow shock region is almost three times higher compared to the $30$ nT case, that is to say, the accumulation of IMF lines leads to a further extension of the bow shock.

\begin{table*}
\centering
\begin{tabular}{c | c c c c} \hline
 & Bow shock & & Magnetopause & \\ \hline
Case ($P_{d}$ nPa,$B_{IMF}$ nT) & $A$ & $B$ & $A$ & $B$ \\ \hline 
$100,30$  & $4.80 \pm 0.06$ & $0.47 \pm 0.02$ & $1.94 \pm 0.12$ & $0.41 \pm 0.03$ \\
$250,30$ & $3.78 \pm 0.08$ & $0.53 \pm 0.02$ & $1.38 \pm 0.06$ & $0.50 \pm 0.02$ \\
$250,100$ & $4.00 \pm 0.11$ & $0.55 \pm 0.03$ & $1.47 \pm 0.10$ & $0.46 \pm 0.03$ \\
\end{tabular}
\caption{Regression parameters ($r_{mp,bs} = A + B R_{ion}$) in simulations with different space weather conditions. The standard errors of the regression parameters are included.}
\label{5}
\end{table*}

In summary, the Earth magnetosphere expansion induced by EUV and X-ray fluxes, hundreds to thousand times larger during the Hadean and Archean eons compared to the present time, leads to an increase of the bow shock and magnetopause standoff distance, reducing the habitability constrains imposed by the SW. On the other hand, the large SW dynamic pressure in the Hadean and Archean eons partially counterbalance the ionosphere expansion, leading to a stronger ionosphere compression that may explain why the standoff distance ratio calculated in the simulations ranges between $0.41$ and $0.55$ for the different space weather conditions analyzed. Thus, any further increase of the SW dynamic pressure causes a stronger compression of the ionosphere, opposing the ionosphere expansion and the increase of the bow shock and magnetopause standoff distance. It should be noted that the IMF lines pile up in the bow shock region causes a further extension of the bow shock standoff distance as the IMF intensity increases while the magnetopause standoff distance decrease, particularly in configurations with Southward IMF orientation. Consequently, the set of simulations performed ignoring the effect of the ionosphere expansion during the Hadean and Archean eons provides a lower bound of the standoff distance, that is to say, the Earth habitability constrains are less restrictive at least for quiet space weather configurations. It should be recalled that, simulations without the effect of the ionosphere expansion for quiet space weather conditions during the Hadean and Archean eons already predict an efficient shielding by the Earth magnetic field. That means, adding the effect of the ionosphere expansion does not change the study conclusions. However, CME-like space weather conditions should be analyzed in more detail and should include the ionosphere expansion to provide a more accurate prediction of the SW $P_{d}$ and $B_{IMF}$ threshold for a direct SW precipitation on the Earth surface. Such analysis will be the topic of future analysis.

\section{Discussion and conclusions}
\label{Conclusions}

The Earth habitability states caused by the solar wind along the Sun's evolution on the main sequence are analyzed considering the evolution of the space weather conditions as well as different Earth magnetic field intensities and topologies inferred from paleomagnetism surveys and numerical models. The study is based on a set of simulations performed with the MHD code PLUTO, calculating the magnetopause standoff distance resulting for different space weather conditions (parameterized by the solar wind dynamic pressure and the intensity of a Southward interplanetary magnetic field) and Earth magnetic field models (intensity and quadrupolar - dipolar coefficient ratios). The data set obtained for the magnetopause standoff distance is fitted with respect to the surface function $log(r_{mp}) = log(Z) + Mlog(|B|_{IMF}) + Nlog(P_{d})$ and compared to the SW dynamic pressure and IMF intensity along the Sun's evolution on the main sequence obtained from the models by \citet{Griemeier,Vidotto,Carolan} and alternatively with \citet{Ahuir} model. The study assumes the Earth habitability is jeopardized if the magnetopause collapses into the Earth surface, that is to say, the solar wind can precipitates directly towards the Earth.

Low field periods with Earth magnetic field intensities below $10$ $\mu$T (Proterozoic eon as well as the Cambrian, Denovian, Carboniferous and the Triassic periods \citet{Elming,Hawkins,Hawkins2,Thallner,Heunemann}) and between $10$-$20$ $\mu$T (Paleo-Archean and Meso-Archean eras, Proterozoic eon, Jurassic and Paleogene periods in the Cenozoic era \citet{Morimoto,Chiara,Prevot,Kosterov,Juarez,Rathert}) are analyzed using dipolar field models with $5$ and $15$ $\mu$T. High field (Hadean eon and the Eo-Archean era \citet{Hale,Tarduo4}) and standard field periods (Meso-Proterozoic and Neo-Archean eras as well as the Neogene and Quaternary periods \citet{Yoshihara,Chiara2,Kaya}) are reproduced by dipolar models with $45$ and $30$ $\mu$T, respectively. In addition, weak multipolar fields linked to geomagnetic reversals (Proterozoic age as well as the Devonian and Cretaceous periods \citet{Lhuillier,Shcherbakova,Guyodo}) are analyzed using Earth magnetic field models with quadrupole - dipole coefficient ratios of $0.2$ and $0.5$ and field intensities of $5$ and $15$ $\mu$T. The Earth magnetic field decay in the Posterum eons is studied using dipolar models with $1$ and $0.1$ $\mu$T, assuming the decline of the geo-dynamo.

The simulations results indicate a dipole with $\geq 5$ $\mu$T can protect the Earth during quiet space weather conditions for all the geological times, except mutipolar models with $5$ $\mu$T and $Q/D \geq 0.2$ during the Hadean eon (including the young sun activity and cyclic modulation). On the other hand, ICME-like space weather conditions require a dipolar magnetic field of $\geq 30$ $\mu$T to protect the Earth from the Archean eon, although a dipolar field with $15$ $\mu$T only provides an efficient shielding from the Mesoarchean era and a dipolar field with $5$ $\mu$T from the Meso-Proterozoic era. It should be noted that the Earth habitability is strongly jeopardized during multipolar low field periods compared to dipole low field periods. Indeed, multipolar field periods with $15$ $\mu$T and $Q/D = 0.2$ cannot provide an efficient magnetic shielding until the Meso-Proterozoic. In addition, multipolar fields with $\leq 5$ $\mu$T and $Q/D \geq 0.2$ are unable to protect the Earth in the Paleozoic era and Posterum eons. Finally, a decaying Earth magnetic field in Posterum eons cannot protect the Earth during ICME events if the dipole intensity decreases below $1$ $\mu$T, or during quiet space weather conditions if the field intensity is $\leq 0.1$ $\mu$T. Table \ref{6} summarizes the geological eras with impacted habitability conditions for each Earth magnetic field configuration analyzed during ICME-like space weather states.

\begin{table*}
\centering
\begin{tabular}{c c | c c c c c c c c} \hline
Eon / Era & Time & $45$ & $30$ & $15$ & ($15$ , $0.2$) & ($15$ , $0.5$) & $5$ & ($5$ , $0.2$) & ($5$ , $0.5$) \\
& (My) & ($\mu$T) & ($\mu$T) & ($\mu$T) & ($\mu$T , $Q/D$) & ($\mu$T , $Q/D$) & ($\mu$T) & ($\mu$T , $Q/D$) & ($\mu$T , $Q/D$) \\ \hline
Hadean & $4400$ & \textcolor{red}{X} & \textcolor{red}{X} & \textcolor{red}{X} & \textcolor{red}{X} & \textcolor{red}{X} & \textcolor{red}{X} & \textcolor{red}{X} & \textcolor{red}{X} \\
Eo-Archean & $3850$ & \textcolor{blue}{O} & \textcolor{blue}{O} & \textcolor{red}{X} & \textcolor{red}{X} & \textcolor{red}{X} & \textcolor{red}{X} & \textcolor{red}{X} & \textcolor{red}{X} \\
Paleo-Archean & $3600$ & \textcolor{blue}{O} & \textcolor{blue}{O} & \textcolor{red}{X} & \textcolor{red}{X} & \textcolor{red}{X} & \textcolor{red}{X} & \textcolor{red}{X} & \textcolor{red}{X} \\
Meso-Archean & $3200$ & \textcolor{blue}{O} & \textcolor{blue}{O} & \textcolor{blue}{O} & \textcolor{red}{X} & \textcolor{red}{X} & \textcolor{red}{X} & \textcolor{red}{X} & \textcolor{red}{X} \\
Paleo-Proterozoic & $1850$ & \textcolor{blue}{O} & \textcolor{blue}{O} & \textcolor{blue}{O} & \textcolor{red}{X} & \textcolor{red}{X} & \textcolor{red}{X} & \textcolor{red}{X} & \textcolor{red}{X} \\
Meso-Proterozoic & $1300$ & \textcolor{blue}{O} & \textcolor{blue}{O} & \textcolor{blue}{O} & \textcolor{blue}{O} & \textcolor{blue}{O} & \textcolor{blue}{O} & \textcolor{red}{X} & \textcolor{red}{X} \\
Neo-Proterozoic & $825$ & \textcolor{blue}{O} & \textcolor{blue}{O} & \textcolor{blue}{O} & \textcolor{blue}{O} & \textcolor{blue}{O} & \textcolor{blue}{O} & \textcolor{red}{X} & \textcolor{red}{X} \\
Paleozoic & $350$ & \textcolor{blue}{O} & \textcolor{blue}{O} & \textcolor{blue}{O} & \textcolor{blue}{O} & \textcolor{blue}{O} & \textcolor{blue}{O} & \textcolor{red}{X} & \textcolor{red}{X} \\
Posterum & $+1500$ & \textcolor{blue}{O} & \textcolor{blue}{O} & \textcolor{blue}{O} & \textcolor{blue}{O} & \textcolor{blue}{O} & \textcolor{blue}{O} & \textcolor{blue}{O} & \textcolor{red}{X} \\
Posterum & $+3000$ & \textcolor{blue}{O} & \textcolor{blue}{O} & \textcolor{blue}{O} & \textcolor{blue}{O} & \textcolor{blue}{O} & \textcolor{blue}{O} & \textcolor{blue}{O} & \textcolor{red}{X} \\
\end{tabular}
\caption{Earth habitability along the Sun's evolution on the main sequence during ICME-like space weather conditions comparing the critical $B_{IMF}$ calculated for different Earth magnetic field configurations with respect to the IMF intensity. A red cross (blue circle) indicates the space weather conditions may hamper (do not hamper) the Earth habitability.}
\label{6}
\end{table*}

The analysis results obtained using different models to predict the space weather conditions during the Sun main sequence are compared, particularly \citet{Griemeier,Vidotto,Carolan} and \citet{Ahuir} models. Some deviations are identified although the main conclusions of the analysis are the same.

The study shows that the Earth atmosphere was exposed to the SW erosion during the Hadean eon, enhancing the water molecules escape as well as the fast Hydrogen and Helium exhaust \citep{Shaw2,Kasting4,Thomassot}. The SW erosion during dipolar low field periods in the Archean eon and Paleo-Proterozoic era may also have an effect on the evolution of the ancient Earth atmosphere. In addition, multipolar low field periods with $\leq 5$ $\mu$T and $Q/D \geq 0.5$ during geomagnetic reversals may lead to an inefficient atmosphere protection in the Phanerozoic eon, identified as a possible driver of climate changes in the Earth by several authors \citep{Verosub,Fairbridge,Schneider,Worm,Valet}.

The implications of our results regarding the SW sterilizing effect on the Earth surface indicate potential effects on the survival and evolution of mats and bacteria from the Paleo-Archean era \citep{Lepot} as well as the multicelular organism from the Meso-Proterozoic era \citep{Butterfield}. Life in the Phanerozoic eon can be also hampered during geomagnetic reversals, linked in several studies with mass-extinction \citep{Kennett,Plotnick,Worm,Isozaki,Wei,Isozaki2}.

The analysis of the ionosphere expansion induced by large EUV and X-ray fluxes in the Hadean and Archean eons, particularity the effect on the bow shock and magnetospause standoff distances, indicates a relaxation of the SW constrains on the Earth habitability requirements. The bow shock and magnetopause standoff distances increase as the ionosphere expands showing an averaged increment ratio of $0.52$ and $0.46$, respectively. The large SW dynamic pressure in the Hadean and Archean eons induces a strong compression of the ionosphere, partially counterbalancing the increment of the bow shock and magnetopause standoff distance as the ionosphere expands. On the other hand, IMF lines pile up in the bow shock region leading to a further extension of the bow shock standoff distance while the magnetopause standoff distance decreases, particularly if the IMF is Southward. Including the ionosphere expansion in the model does not change the analysis results for quite space weather conditions during the Hadean and Archean eons, because the simulation set performed ignoring this effect already indicates the Earth magnetic field provided an efficient protection. Nevertheless, CME-like space weather conditions must be studied including the expansion of the ionosphere to improve the analysis accuracy.

We conclude that the hamper of space weather on the Earth habitability must be considered as an important driver of the atmosphere and life evolution. It must be recalled that there are other important external actuators not included in the study such us impacts events as Geo-crossing asteroids or comets, gamma ray burst, supernovas (cosmic rays) as well as the ultraviolet and X rays radiation generated by the Sun. Impacting events produce dust and aerosols that can cover the Earth atmosphere during tens to hundred of years inhibiting photosynthesis \citep{Alvarez}. An impacting event was the most probable driver of the Cretaceous–Paleogene extinction event \citep{Schulte}. Gamma ray burst may cause the destruction of the Earth's ozone layer that brings protection against the ultraviolet radiation and could be responsible of the late Ordovician mass extinction \citep{Melott,Thomas}. Supernova radiation produces strong ionization effects and induce mass extinction events too \citep{Whitmire,Ellis,Dar,Svensmark}. Likewise, ultraviolet and X rays radiation from the Sun, particularly at the early phases of the Sun evolution, may have an harmful effect on biological structures and could play a role in life evolution \citep{Cockell2,Dartnell,Lammer2,Kaltenegger,Airapetian5}. On top of that, high X-ray and EUV fluxes generated by the young Sun will also be responsible for high neutral loss rates of the atmosphere \citep{Tian,Tian2,Johnstone3}. Nevertheless, the atmospheric erosion and surface sterilization caused by the SW should be included in the list of the most important external effects in considering solar-type Earth and exoplanets habitability along the Sun and other stars evolution.

\section*{Acknowledgements}

This work was supported by the project 2019-T1/AMB-13648 founded by the Comunidad de Madrid. The research leading to these results has received funding from the grants ERC WholeSun 810218, Exoplanets A, INSU/PNP. P. Zarka acknowledges funding from the ERC No $101020459$—Exoradio. We extend our thanks to CNES for Solar Orbiter, PLATO and Meteo Space science support and to INSU/PNST for their financial support. This work has been supported by Comunidad de Madrid (Spain) - multiannual agreement with UC3M (“Excelencia para el Profesorado Universitario” - EPUC3M14 ) - Fifth regional research plan 2016-2020. Data available on request from the authors.

\section*{Data Availability}
 
4.3 version of the PLUTO code used for performing the simulations is available via http://plutocode.ph.unito.it and developed openly at the Dipartimento di Fisica, Torino University in a joint collaboration with INAF, Osservatorio Astronomico di Torino and the SCAI Department of CINECA (Mignone et al., 2007).



\bibliographystyle{mnras}
\bibliography{References} 




\appendix

\section{Extended ionosphere model}

The ionosphere model is upgraded to evaluate the effect of the ionosphere expansion on the bow shock and magnetopause standoff distance. The ionospheric domain is now divided in two sections. At $R = 2.3 - 2.5 R_{E}$ it is defined the coupling region between the upper atmosphere and the ionosphere. From $R = 2.5 R_{E}$ to the ionosphere upper limit (an input of the model) is defined the upper ionosphere region.

Special conditions are imposed in the coupling region between the upper atmosphere and the ionosphere. The plasma density is stratified with respect to the Alfv\'{e}n velocity and the module of the Alfv\'en velocity is fixed ($\mathrm{v}_{A} = 2 \cdot 10^{4}$ km/s) and does not evolve during the simulation, defined as
\begin{equation}
\label{eqn:1}
\rho = \frac{|B|^{2}}{\mu_{0}v_{A}^{2}}.
\end{equation}

The hydro-static equilibrium between the upper atmosphere and the ionosphere is set by fixing the plasma pressure with respect to the sound speed of the SW ($c_{sw}$) and the upper atmosphere ($c_{p}$):
\begin{equation}
\label{eqn:2}
p = \frac{n}{\gamma} \left( \frac{(c_{p} - c_{sw})(r^{3}-R_{s}^{3})}{R_{un}^{3}-R_{s}^{3}} + c_{sw} \right)^{2},
\end{equation}
with $c_{p} = \sqrt{\gamma K_{B} T_{p}/m_{p}}$ with $T_{p}$ the plasma temperature at the upper atmosphere and $T_{sw}$ the SW temperature.

The plasma density and magnetic field are free to evolve in the upper ionosphere region, thus the upper ionosphere can be compressed by the effect of the SW as well as the Earth magnetic field lines. Only the slope of the pressure is constrained following equation \ref{eqn:2}.

The field-aligned currents ($J_{FAC}$) are calculated in the upper ionosphere and the coupling region between the upper atmosphere and the ionosphere based on \citep{Buchner} model:
\begin{equation}
\label{eqn:3}
\mathbf{J}_{FAC} = \mathbf{J} - \mathbf{J}_{\perp},
\end{equation}
where
\begin{equation}
\label{eqn:4}
\mathbf{J} = \frac{1}{\mu_{0}} \mathbf{\nabla} \times \mathbf{B}
\end{equation}
\begin{equation}
\label{eqn:5}
\mathbf{J_{\perp}} = \mathbf{J} - \frac{J_{r} B_{r} + J_{\theta} B_{\theta} + J_{\phi} B_{\phi}}{|B|^{2}} \mathbf{B},
\end{equation}
with $\mathbf{J}$ the plasma current, $\mathbf{J}_{\perp}$ the perpendicular component of the plasma current along the magnetic field line, $\mu_{0}$ the vacuum magnetic permeability, and $\mathbf{B}$ the magnetic field.

Next, the electric field is calculated using the Pedersen conductance ($\sigma$) empirical formula,
\begin{equation}
\label{eqn:6}
\sigma = \frac{40 E_{0} \sqrt{F_{E}}}{16 + E_{0}^{2}},
\end{equation}
with $E_{0} = K_{B} T_{e}$ the mean energy of the electrons, $F_{E} = n_{e} \sqrt{E_{0}/(2 \pi m_{e})}$ the energy flux, and $K_{B}$ the Boltzmann constant ($T_{e}$ and $m_{e}$ are the electron temperature and mass, respectively). Thus, the electric field ($E$) linked to the FAC is:
\begin{equation}
\label{eqn:7}
\mathbf{E} = \sigma \mathbf{J}_{FAC},
.\end{equation}
When the electric field is calculated, the velocity of the plasma in the coupling region between the upper atmosphere and the ionosphere is:
\begin{equation}
\label{eqn:8}
\mathbf{v} = \frac{\mathbf{E} \times \mathbf{B}}{|B|^{2}}.
\end{equation}



\bsp	
\label{lastpage}
\end{document}